\documentclass[%
 reprint,
 amsmath,amssymb,
 aps,
]{revtex4-2}

\usepackage{graphicx}% Include figure files
\usepackage{dcolumn}% Align table columns on decimal point
\usepackage{bm}% bold math
\usepackage{comment}

\newcommand{\la}{\langle}
\newcommand{\ra}{\rangle}

\usepackage{xcolor}
\newcommand{\ps}[1]{\textcolor{black}{#1}}

\begin{document}

\preprint{APS/123-QED}

%\title{The Universal Structure of Turbulent Radiative Mixing Layers}
\title{\ps{ Structure of Turbulent Radiative Mixing Layers}}

\author{Prateek Sharma$^{1}$}
\email{prateek@iisc.ac.in}
\author{Arnav Kumar$^{1}$}
\author{Dipayan Datta$^{1}$}
\author{Arif Babul$^{1,2}$}
%\altaffiliation{Also at Institute for Astronomy, University of Edinburgh, Blackford Hill, Edinburgh EH9 3HJ, UK}
\author{Rishita Das$^{3}$}
\author{Konduri Aditya$^{4}$}

\affiliation{$^{1}$Department of Physics, Indian Institute of Science, Bangalore 560012, India}
\affiliation{$^{2}$Department of Physics and Astronomy, University of Victoria, Victoria, BC V8P 1A1, Canada}
\affiliation{$^{3}$Department of Aerospace Engineering, Indian Institute of Science, Bangalore 560012, India}
\affiliation{$^{4}$Department of Computational and Data Sciences, Indian Institute of Science, Bangalore 560012, India}

\date{\today}

\begin{abstract}
Turbulent radiative mixing layers (TRMLs), where shear-driven turbulence mixes dense and diffuse gas into rapidly cooling intermediate-temperature gas, are ubiquitous in the interstellar and circumgalactic media. A Reynolds decomposition of 3D simulations shows that a cooling-driven compressive stress creates a pressure dip that sustains a steady inflow of hot gas, and that radiative losses are balanced mainly by enthalpy advection, not turbulent heat flux. Contrary to previous claims, the cooling rate per unit area saturates and becomes independent of box size in the strong-cooling limit.
\end{abstract}

\maketitle

%\tableofcontents

%\section{\label{sec:introduction} Introduction}

\emph{Introduction}--- Turbulent shear flows appear in engineering, geophysical, and astrophysical contexts~\cite{Bradshaw1975,Baklanov2011,Begelman1990}. Their adiabatic evolution is well studied~\cite{Dimotakis2005,Smits2006}, while combustion adds complexity through heat release~\cite[Ch.~14]{Landau1987}. The astrophysical analogue is the turbulent radiative mixing layer (TRML)—a shear interface where intermediate-temperature plasma cools rapidly~\cite{Begelman1990,Kim2013,Fraternali2017}.

TRMLs are the simplest realizations of multiphase radiative turbulence. They form the building blocks of radiative cloud-crushing~\cite{Armillotta2016,Gronke2018,Kanjilal2021MNRAS}, cosmological cold filament accretion~\cite{Mandelker2020}, and the interaction of galactic winds with the circumgalactic medium (CGM)~\cite{Soklowska2018}. Observations reveal multiphase gas spanning \(10^4\)--\(10^7\) K in the CGM, consistent with the ubiquity of such layers~\cite{Tumlinson2017review}. 

Yet a description of their turbulent global structure has remained elusive. Here \ps{we carry out a Reynolds decomposition of TRMLs in statistical steady-state and} show that a positive vertical compressive stress ${\cal R}_{zz}$, driven by radiative cooling and typically negligible relative to pressure in non-radiative shear layers, causes the TRML pressure to dip, allowing the hot phase to be sucked into the cold interface. \ps{We quantify the energy balance in TRMLs and show that in the strong cooling regime, their cooling rate densities reach a value independent of the box-size.} 

Non-radiative shear layers obey a similarity solution: the mixing layer grows linearly $\propto \Delta u t$ ($\Delta u$ is the imposed shear velocity; see Section~S1 in Supplemental Material (SM) for the governing equations and parameters), and expands preferentially into the diffuse medium~\cite{Barenblatt1996,Dimotakis1986}; \ps{see section S2 in SM}. Radiative losses fundamentally alter this picture: TRMLs do not grow indefinitely but reach a quasi-steady turbulent state.

\emph{Turbulent Radiative Mixing Layers}--- At early times, a radiative shear layer evolves similarly to its adiabatic counterpart. After a cooling time, however, the system ceases to grow and settles into a statistically steady turbulent radiative mixing layer (TRML). A fiducial cooling time \ps{(the ratio of thermal energy and energy loss rate)} is
\begin{equation}
\label{eq:t0}
t_0 = t_{\rm cool}(T_0 = \sqrt{T_c T_h}) = 
\frac{p_0}{(\gamma - 1) n_0^2 \Lambda(T_0)},
\end{equation}
where \(T_0\) is the geometric mean of the cold and hot temperatures ($T_c$, $T_h$; $n_0$ and $p_0$ are the corresponding number density and pressure), \(\Lambda(T)\) the cooling function \cite{Schure2009} \ps{that quantifies the temperature dependence of cooling losses, and $\gamma=5/3$ the adiabatic index (see Fig. 3 in SM)}. After a few $t_0$, horizontally and ensemble-averaged quantities satisfy \(\partial / \partial t = 0\).

\begin{figure}[t]
\includegraphics[scale=0.6]{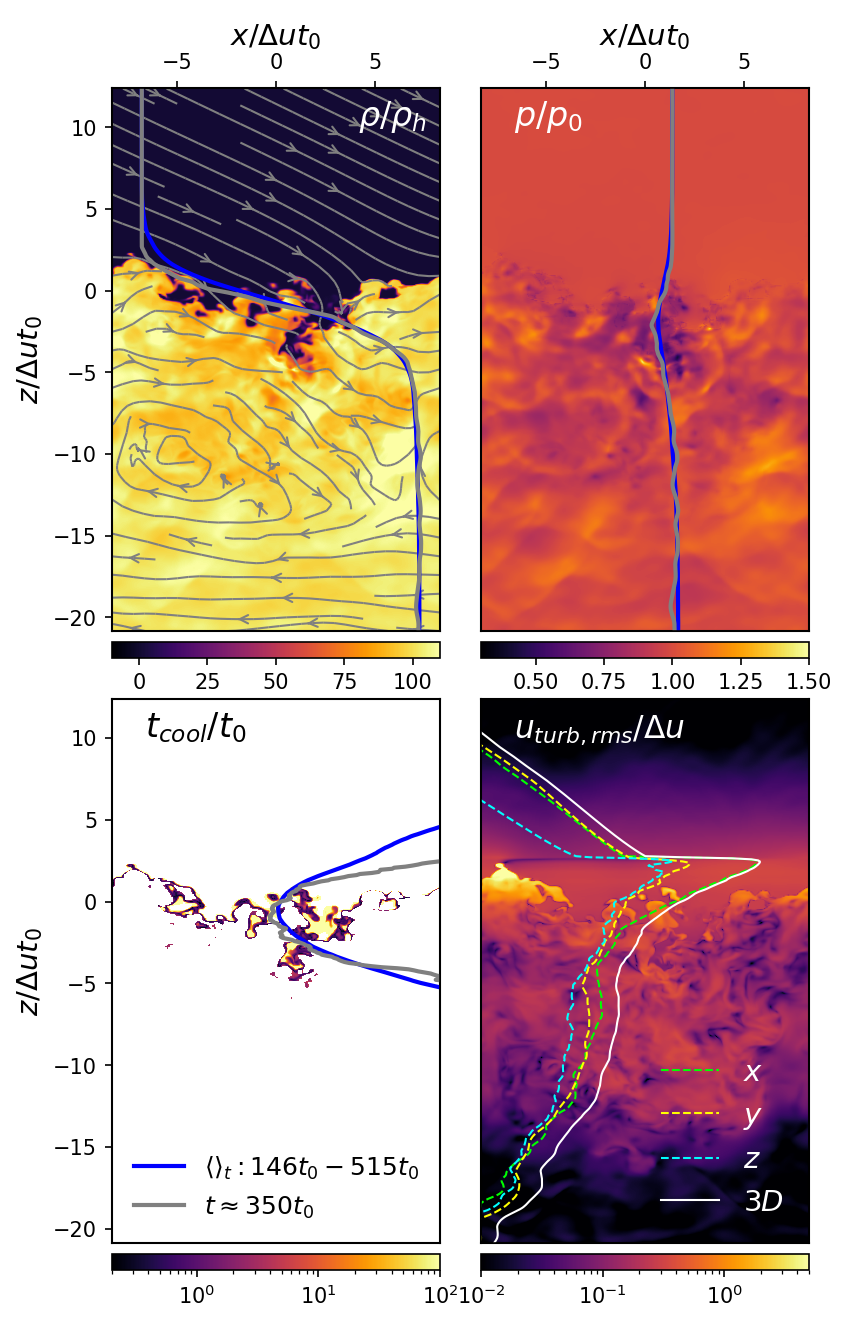}
\caption{\label{fig:snapshots}
Snapshots showing density, pressure, cooling time (Eq. \ref{eq:t0} evaluated at local $n,~T$), and velocity fluctuations (mass-weighted) in the quasi-steady TRML. Diffuse hot gas collapses quasi-isobarically through a thin fractal cooling interface. \ps{Streamlines in the density panel show the instantaneous velocity field. The blue and gray profiles in respective panels show the time-averaged and instantaneous horizontally-averaged profiles.}
}
\end{figure}

In the absence of \ps{shear-driven} turbulence, steady subsonic cooling flows are possible only in cylindrical or spherical geometries~\cite{Dutta2022}. Steady isobaric solutions with conduction~\cite{Kim2013,Tan2021} show a monotonically decreasing velocity toward the cold phase. Simulations show that plane-parallel TRMLs also reach turbulent quasi-steady states~\cite{Ji2019,Fielding2020ApJ}, suggesting that turbulent correlations act as an effective conductive process in radiative boundary layers.\footnote{We have verified that in the absence of imposed shear ($\Delta u = 0$), the two phases remain sharply separated and $\dot{\Sigma}_{\rm cool} \approx 0$ in steady state. {\ps{Unlike a TRML, there is no natural cooling length scale with $\Delta u = 0$. The temperature across a TRML shows large fluctuations, and simple diffusion does not describe the local turbulent heat transport.}}} But deviations from isobaricity tightly couple the momentum and energy %evolution. 
\ps{transport}. Analytic models that include turbulent viscosity and conductivity~\cite{Chen2023} or entrainment scalings~\cite{Ji2019,Tan2021} reproduce some aspects of TRML dynamics, but disagree on the origin of hot-gas inflow—whether driven by a pressure deficit~\cite{Ji2019} or by turbulent stresses~\cite{Fielding2020ApJ,Tan2021,Chen2023}. However, missing the compressive ${\cal R}_{zz}$ stress \ps{highlighted in our work}, the existing analytic models fail to reproduce the prominent pressure dip observed in simulations (Fig.~\ref{fig:snapshots}). 

We present a quasi-steady state analysis in the TRML rest frame, based on Reynolds decomposition into mean and turbulent components \cite{Reynolds1894}. While turbulent shear stresses are well understood in incompressible shear flows, we show that vertical compressive stress in TRMLs due to strong compressibility driven by cooling fundamentally alters their structure relative to their non-radiative counterparts. In steady state, hot diffuse gas is compressed by factors of \(\sim 100\) as it mixes, cools, and settles into the cold phase.

We test our framework with 3D hydrodynamic simulations using the \textsc{AthenaK} code~\cite{Stone2024}. The setup places cold dense gas below hot diffuse gas, in pressure equilibrium but with a shear \(\Delta u\). To maintain approximate TRML stationarity, the cold-phase \ps{shear} velocity ($u_x$) is set to $-\chi^{-1/2}$ \ps{($\chi$ is the cold-hot density ratio)} times that of the hot phase~\cite{Dimotakis1986}. The interface is initialized with a smoothed $\tanh$ profile \ps{ in $z$ for $u_x$ and density, with \ps{vertical velocity} $u_z$ in the transition region} seeded with small white-noise perturbations. \ps{All other quantities are initialized to a constant value}. To isolate the TRML dynamics, cooling is turned off in the two stable phases \(T<1.05\,T_c\) and \(T>0.95\,T_h\). Boundaries are periodic in \(x,y\), reflective at the bottom, and inflow/outflow at the top with the initial shear imposed. The domain for our fiducial run spans \([10,10,40]\) pc in \([x,y,z]\) ($x$ [$y$, $z$] is the streamwise [spanwise, shear] direction). While we focus on this run in the Letter, we have carried out several runs with varying parameters (resolution, box-size, $\rho_h$, $\Delta u$) presented in Table 1 of SM. \ps{Section 6 in SM presents results from widely varying `engineered' cooling functions.} 

The Kelvin–Helmholtz (KH) instability develops at the interface, and the system reaches a turbulent steady state after \(\sim t_0\). Figure~\ref{fig:snapshots} shows $x$--$z$ slices of density, pressure, cooling time, and rms velocity fluctuations. Diffuse hot bubbles are wrapped in KH rolls and collapse quasi-isobarically, driving acoustic waves into the cold phase. Velocity fluctuations are highly anisotropic, dominated by the streamwise velocity fluctuations. Although the thermal boundary layer is comparatively broad, the cooling interface is thin and irregular, a robust hallmark of radiative mixing layers \cite{Fielding2020ApJ,Tan2021}. 

\begin{figure}[b]
\includegraphics[scale=0.42]{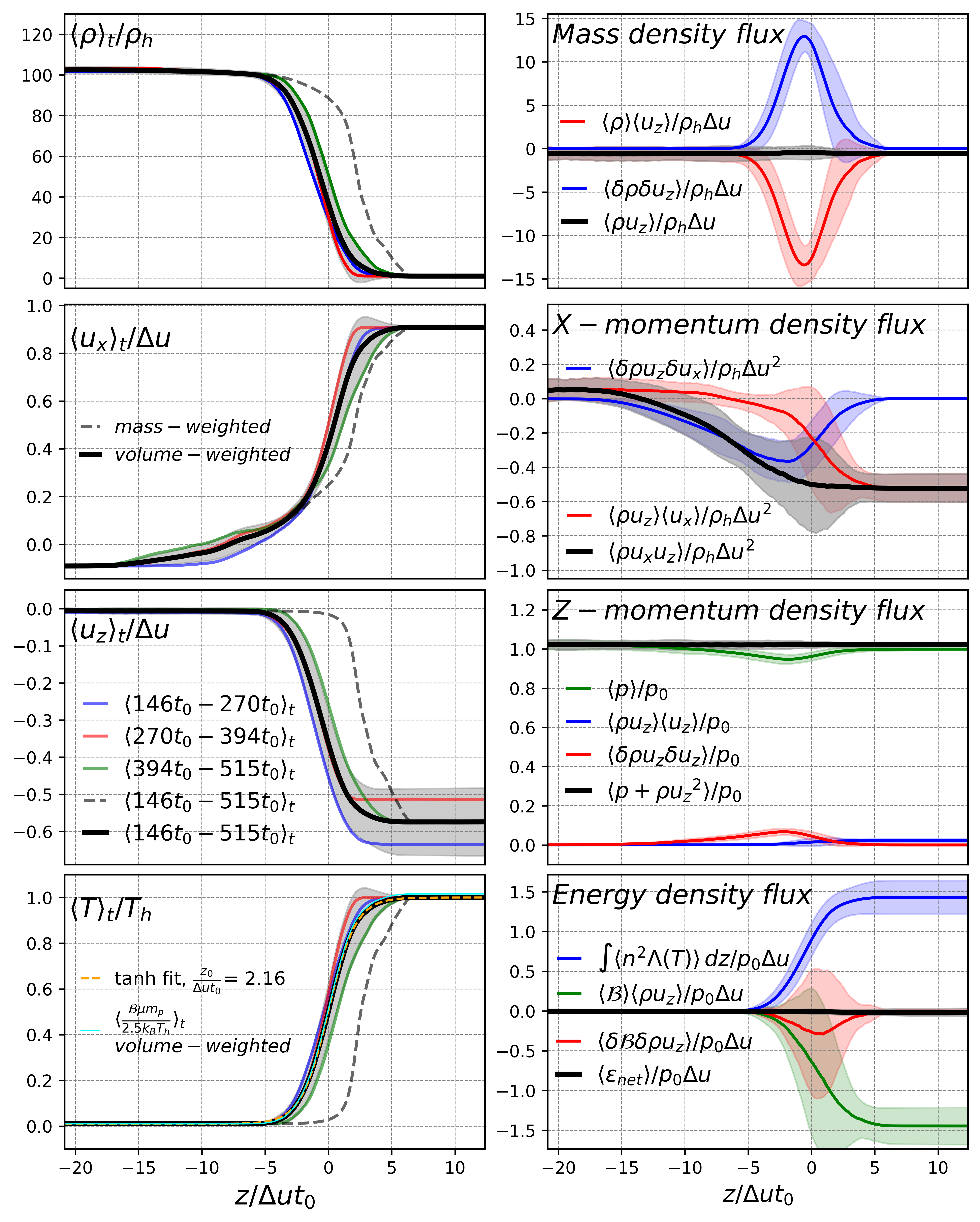}
\caption{\label{fig:mean_fluxes}
Mean vertical profiles of density, velocity components, temperature \ps{(left panels)}, and %conserved 
fluxes \ps{(right panels)} in steady state. Shaded bands indicate $1\sigma$ temporal variability. Turbulent contributions to mass and energy fluxes are confined to the TRML, while the x-momentum flux progressively penetrates into the cold phase. The temperature profile is well fit by a tanh function. 
 }
\end{figure}

\emph{Reynolds Decomposition in Steady State}--- In the steady state, the mean density profile is stationary in the TRML frame, which slowly drifts upward in the simulation frame, in which the cold phase is stationary due to a reflecting bottom boundary. The vertical velocity of the TRML is obtained by enforcing $\partial_t \langle \rho \rangle = 0$ in the TRML frame (here $\langle \cdot \rangle$ denotes horizontal and temporal averaging in quasi-steady state) and integrating the continuity equation across the turbulent interface,
\[
v_{{\rm TRML}_z} = \frac{(\rho u_z)_h - (\rho u_z)_c}{\rho_h - \rho_c}.
\]
 In the simulation frame, this reduces to $v_{{\rm TRML}_z} = -u_{z,h}/(\chi-1)$ \ps{($u_{z,h}$ is the mean vertical velocity in the upstream hot phase)}. Unless stated otherwise, velocities are henceforth measured in the TRML frame in which the horizontally-averaged density is stationary (top-left panel of Fig.~\ref{fig:mean_fluxes}).

Once steady state is established, the net vertical mass flux $\langle \rho u_z \rangle$ is uniform \ps{and non-zero} across the layer (top-right panel of Fig.~\ref{fig:mean_fluxes}). Within the TRML, rising dense plumes and sinking hot clumps generate a strong positive turbulent mass flux $\langle \delta \rho \, \delta u_z \rangle$ (top-right panel).

In contrast, the streamwise ($x$-direction) momentum does not equilibrate. Its mean evolution obeys
\begin{equation}
\label{eq:mean_u_x}
\frac{\partial}{\partial t} \langle \rho u_x \rangle + \frac{\partial}{\partial z} \langle \rho u_x u_z \rangle = 0,
\end{equation}
with the flux decomposed as $\langle \rho u_x u_z \rangle = \langle \rho u_z \rangle \langle u_x \rangle + {\cal R}_{zx}$, where ${\cal R}_{zx} = \langle \delta (\rho u_z) \, \delta u_x \rangle < 0$ is the Reynolds shear stress.\footnote{\ps{Our definition of the stress tensor differs from the usual one in which ${\cal R}$ is a symmetric tensor.}} Although $\partial_t \langle \rho \rangle = - \partial_z \langle \rho u_z \rangle = 0$ in the TRML frame, Eq.~\eqref{eq:mean_u_x} is Galilean invariant along $x$, so a uniform boost in the $x$-direction cannot enforce $\partial_t \langle u_x \rangle = 0$.

Integrating Eq. \ref{eq:mean_u_x} vertically across the turbulent layer yields
\begin{equation}
\label{eq:TRML_px}
\frac{d}{dt} \int \langle \rho u_x \rangle \, dz = - \langle \rho u_z \rangle \Delta u,
\end{equation}
where $\Delta u = u_{x,h} - u_{x,c}$. Thus, the total horizontal momentum in the mixing layer increases as high-$u_x$ gas enters from the hot phase and an equal mass of low-$u_x$ gas exits into the cold phase. As shown in panel (2,1) of Fig.~\ref{fig:mean_fluxes} (second row, first column), $\langle u_x \rangle$ saturates inside the TRML while its penetration into the cold phase increases slowly over time. This lack of strict steady state in $x$-momentum is consistent with earlier numerical findings~\cite{Chen2023}.

The vertical momentum equation does attain steady-state:
\begin{equation}
\label{eq:mean_pz}
\frac{d}{dz} \left[ \langle \rho u_z \rangle \langle u_z \rangle + {\cal R}_{zz} + \langle p \rangle \right] = 0,
\end{equation}
where ${\cal R}_{zz} = \langle \delta (\rho u_z) \, \delta u_z \rangle$ is the vertical turbulent compressive stress. In incompressible shear flows, the analogous normal stress produces a well-known mean-pressure deficit, $\langle p\rangle + \rho\langle \delta u_z^2\rangle \approx$ const~\cite{Tennekes1972}; in subsonic non-radiative mixing layers, however, it is negligible relative to the thermal pressure (${\cal R}_{zz}/p \sim 10^{-3}$; Section S2 of SM). In TRMLs, cooling-driven compression raises it to ${\cal R}_{zz} \approx \rho_h \Delta u^2$ (${\cal R}_{zz}/p \approx 0.07$), producing a pressure dip that sustains a steady inflow of hot gas. This term has not been included in prior analytic models of TRMLs.

The total vertical momentum flux—the sum of advective, turbulent, and thermal pressures—is conserved (panel [3,2] of Fig.~\ref{fig:mean_fluxes}). A dip in $\langle p \rangle$ forms within the TRML, supported by ${\cal R}_{zz}>0$ from rising dense plumes and descending diffuse bubbles. The sum of thermal and advective pressure \ps{($\langle \rho u_z\rangle \langle u_z \rangle$)} in the upstream balances the downstream thermal pressure, resulting in a density slightly larger than $\chi \rho_h$ in the downstream.

The steady-state energy equation reads
\begin{equation}
\label{eq:total_energy}
\frac{d}{dz} \langle \rho u_z {\cal B} \rangle = - \langle n^2 \Lambda(T) \rangle,
\end{equation}
with Bernoulli number ${\cal B} = u^2/2 + \gamma p / [(\gamma-1)\rho]$, dominated by enthalpy in the subsonic regime we consider.\footnote{\ps{ Because of multiphase turbulence in the TRML, $\langle n^2 \Lambda(T) \rangle \neq \langle n \rangle^2 \Lambda (\langle T \rangle )$.}} Integrating across the TRML shows that radiative losses in TRML are balanced by the divergence of vertical energy flux.

Viscous and adiabatic heating merely redistribute energy between kinetic and thermal forms and are omitted from Eq.~\eqref{eq:total_energy}. Locally, cooling is balanced by the divergence of (i) enthalpy flux, $\langle {\cal B} \rangle \langle \rho u_z \rangle$ (i.e. $p dV$ work), and (ii) turbulent heat flux ${\cal Q}_t = \langle \delta {\cal B} \, \delta (\rho u_z) \rangle$. Panel (4,2) of Fig.~\ref{fig:mean_fluxes} shows this partition: cumulative cooling losses (blue) are offset by enthalpy advection (green) and turbulent flux (red). Similar global energy balance, where isobaric heating offsets cooling, has been invoked for cold-gas condensation in cluster cores~\cite{banerjee2014turbulence,Mohapatra2022characterising}.

\begin{figure}[b]
\includegraphics[scale=1.1]{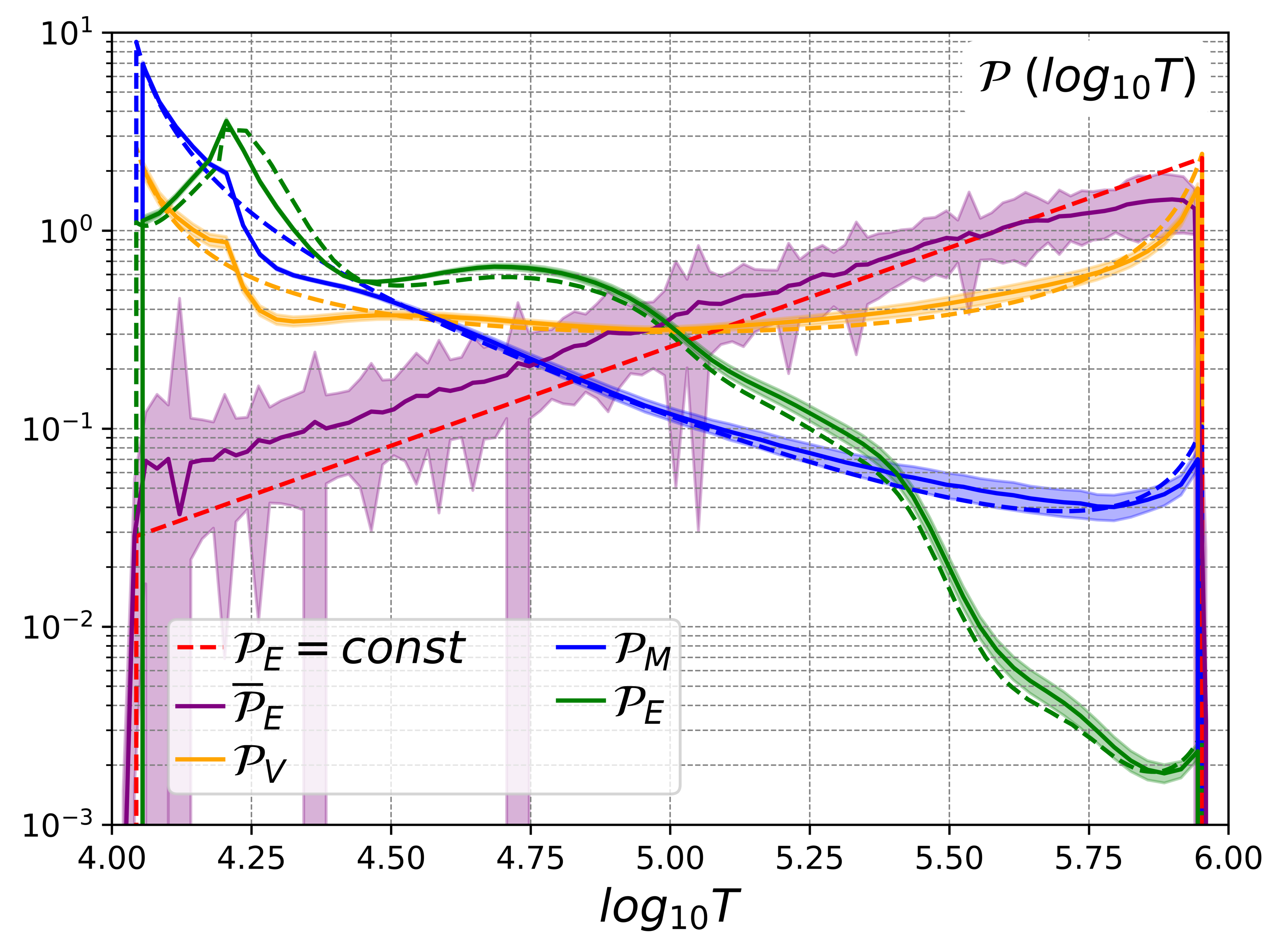}
\caption{Normalized volume, mass, and emissivity PDFs (${\cal P}[\log_{10}T]$) across temperature within the TRML, restricted to $[1.1T_c,0.9T_h]$ where the mean temperature profile is well described by Eq.~\eqref{eq:tanh_T}. Solid lines show simulation results (shaded bands: temporal $1\sigma$ variations); dashed lines show analytic predictions based on Eq. \ref{eq:PV_to_PMPE} and the $\tanh$ fit to ${\overline{\cal P}}_V(\langle T \rangle)$, which remarkably also fits ${\cal P}_V(T)$.\footnote{\ps{$\overline{\cal P}_V \simeq {\cal{P}}_V$ is not generic but holds only for our cooling function. However, the isobaricity assumption leading to Eq. \ref{eq:PV_to_PMPE} and tanh fit to $\langle T \rangle$ hold more broadly (tested at $\xi \approx 2.6$), as elaborated using `engineered' cooling functions in section S6 of SM.}} In the turbulent steady state, ${\cal P}_{V,M,E}(T)$ show tiny variations in time for all the resolved runs. For comparison, the red dashed line shows the distinct prediction for ${\cal P}_E$ from a non-turbulent, steady cooling flow. The purple line denotes the emissivity PDF across horizontally averaged temperature, distinct from that across physical temperature. \label{fig:PDFs}}
\end{figure}

\emph{Thermal structure of TRMLs}--- The bottom-left panel of Fig.~\ref{fig:mean_fluxes} shows that the {\em mean} \ps{(horizontally-averaged)} TRML temperature profile is well described by
\begin{equation}
\label{eq:tanh_T}
\langle T \rangle(z) = \frac{T_h-T_c}{2}\tanh\!\left(\frac{z}{z_0}\right)+\frac{T_h+T_c}{2},
\end{equation}
with $z=0$ defined at $\langle T \rangle=(T_h+T_c)/2$. This one-to-one mapping between $z$ and $\langle T \rangle$ enables an analytic prediction for the volume PDF of $\langle T \rangle$, 
\[
\overline{{\cal P}}_V(\langle T \rangle)\,d\langle T \rangle \equiv \frac{dz}{\Delta z}
= \frac{d\langle T \rangle}{\Delta z\,T'},
\]
where $T' = d\langle T \rangle/dz$ and $\Delta z$ is the extent over which Eq.~\eqref{eq:tanh_T} is a good fit ($1.1T_c<\langle T \rangle<0.9T_h$).  
Using the inverse of Eq.~\eqref{eq:tanh_T}, the gradient is
\[
T' = \frac{T_h-T_c}{2z_0}\left[1-4\left(\frac{\langle T \rangle-(T_h+T_c)/2}{T_h-T_c}\right)^2\right].
\]

In quasi-isobaric TRMLs the mass- and emissivity-weighted PDFs \ps{(relative to the physical temperature $T$, rather than the mean temperature $\langle T \rangle$ above)} follow
\begin{equation}
\label{eq:PV_to_PMPE}
{\cal P}_M(T)\propto\frac{{\cal P}_V(T)}{T}, \qquad
{\cal P}_E(T)\propto\frac{{\cal P}_V(T)\Lambda(T)}{T^2},
\end{equation}
in excellent agreement with simulations (compare the solid and dashed lines for ${\cal P}_{V,M,E}(T)$ in Fig.~\ref{fig:PDFs}). The emissivity PDF (${\cal P}_E(T)$) differs qualitatively from the steady cooling flow prediction (green lines versus red dashed line; \cite{Dutta2022}).  In contrast to TRMLs, for a non-turbulent cooling flow, the physical and average temperatures are the same.

The link to energy transport becomes clear by rewriting the TRML energy equation (Eq. \ref{eq:total_energy}) as
\begin{equation}
\label{eq:energy_diff}
\dot{\Sigma}_M \frac{d}{dz}\langle {\cal B} \rangle - \frac{d{\cal Q}_t}{dz}
= T' \, \dot{\Sigma}_{\rm cool}\,\overline{{\cal P}}_E(\langle T \rangle)
= \langle n^2\Lambda\rangle,
\end{equation}
where $\dot{\Sigma}_M=-\langle \rho u_z\rangle$ is the entrainment rate, ${\cal Q}_t=\langle \delta{\cal B}\,\delta(\rho u_z)\rangle$ is the turbulent energy flux, and $\dot{\Sigma}_{\rm cool}$ is the net cooling rate per unit area.  This relation links the emissivity PDF, turbulent flux, and entrainment rate. Global energy balance requires $\dot{\Sigma}_M\Delta{\cal B}=\dot{\Sigma}_{\rm cool}$, with $\Delta{\cal B}$ the jump in Bernoulli parameter across the TRML. Integrating Eq.~\eqref{eq:energy_diff} yields
\begin{equation}
\label{eq:Q_t_closure_int}
\frac{{\cal Q}_t(\langle T \rangle)}{\dot{\Sigma}_{\rm cool}}
\approx -\!\left[\overline{{\cal C}}_E(\langle T \rangle)-\frac{\langle T \rangle-T_c}{T_h-T_c}\right],
\end{equation}
where $\overline{{\cal C}}_E(\langle T \rangle)=\int_{T_c}^{\langle T \rangle}\overline{{\cal P}}_E(s)\,ds$ is the cumulative emissivity distribution. For subsonic TRMLs ${\cal B}\simeq (5/2)k_BT/\mu m_p$ (see bottom-left panel of Fig. \ref{fig:mean_fluxes}).  

\begin{figure}[b]
\includegraphics[scale=1.1]{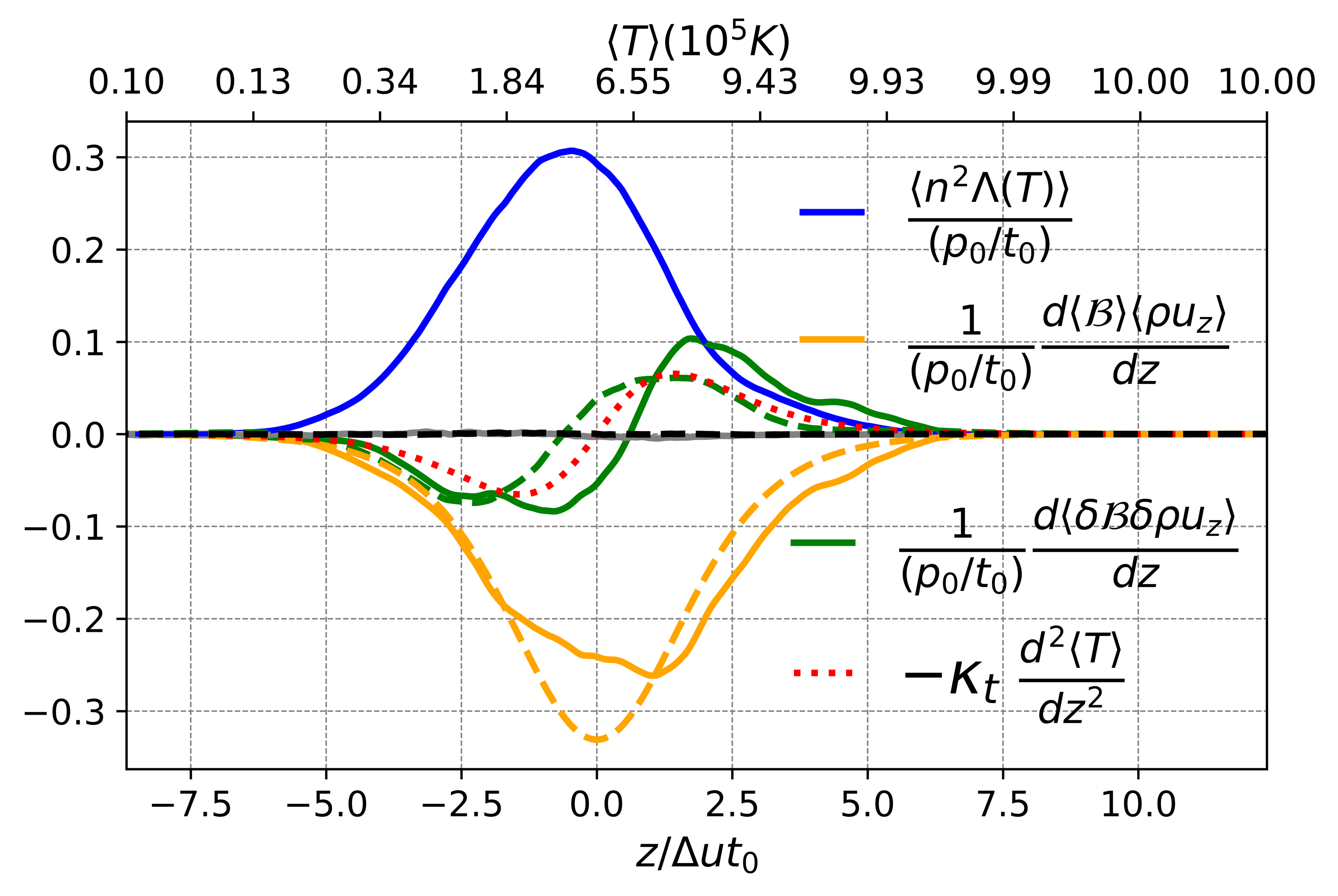}
\caption{The balance of radiative cooling losses $\la n^2 \Lambda \ra$, divergence of enthalpy flux ($d[\la {\cal B} \ra \la \rho u_z \ra]/dz$), and turbulent heat transport ($d[\la \delta {\cal B} \delta \rho u_z \ra]/dz$)  within the TRML. Solid lines are based on simulation data and dashed lines on the theoretical model (Eqs. \ref{eq:tanh_T}, \ref{eq:Q_t_closure_int}). The red dotted line for turbulent heating shows a simple closure ${\cal Q}_t = - \kappa_t d\la T \ra/dz$ with $\kappa_t = 0.1 k_B \rho_h \Delta u^2 t_0 /(\mu m_p)$. \label{fig:thermal_balance}}
\end{figure}

The turbulent flux peaks at $T^*$ such that $\overline{{\cal P}}_E(T^*)\simeq 1/\Delta T$ ($\Delta T = T_h - T_c$), the flat emissivity distribution expected in a pure cooling flow. Thus, once $\dot{\Sigma}_{\rm cool}$, $\langle T \rangle$, and $\overline{{\cal P}}_E$ are calibrated, other macroscopic quantities such as ${\cal Q}_t$, $\langle \rho \rangle$, and $\langle u_z \rangle$ follow directly. The solid purple line in Fig. \ref{fig:PDFs} shows the emissivity PDF relative to the mean temperature, which is qualitatively different from the PDF relative to the physical temperature and slightly larger than the cooling flow prediction at low temperatures $\langle T \rangle$ because of heating due to turbulent heat transport. 

Fig. \ref{fig:thermal_balance} shows that turbulence transports thermal energy from larger $z$ to smaller (on average, hotter to cooler regions of TRML). In striking contrast to analytic 1D models \cite{Kim2013,Tan2021Model,Chen2023}, heating/cooling due to the divergence of turbulent heat flux is subdominant at most $z$, with the dominant balance between cooling and the divergence of enthalpy flux. Most interestingly, the emergent radiative loss rate density $\la n^2 \Lambda \ra$ (blue solid line) is qualitatively different in shape from  $\Lambda(T)/T^2$ expected in the absence of multiphase turbulence.

\emph{Discussion}--- Shear-driven turbulence fundamentally changes the radiative cooling layers between dense and diffuse CGM phases. Crucially, the pressure dip caused by the compressive ${\cal R}_{zz} > 0$ stress driven by cooling allows a steady, subsonic cooling flow. Unlike shocks—whose structure is fixed entirely by conservation laws—TRML structure emerges from turbulence–cooling coupling. These results enable numerical calibration of the entrainment rate as a function of key parameters of the setup (${\cal M}$, ${\chi}$, $\rho_h$, $\Delta u$) \cite{Fielding2020ApJ}. For a set of runs with different parameters, including varying box-size and resolution, we have quantified the turbulent transport coefficients $\la \delta \rho \, \delta u_z \ra$, ${\cal R}_{zx}$, ${\cal R}_{zz}$, and ${\cal Q}_t$ (Table 1 in SM), which may lead to subgrid prescriptions for TRMLs in cosmological simulations \cite{Ramesh2024MNRAS}. A generalization of our approach to include important physical effects such as magnetic fields \cite{Zhao2023,Das2024} and thermal conduction \cite{armillotta2017}, via Maxwell stress and microscopic transport, is a natural next step.

\begin{figure}
\includegraphics[scale=0.75]{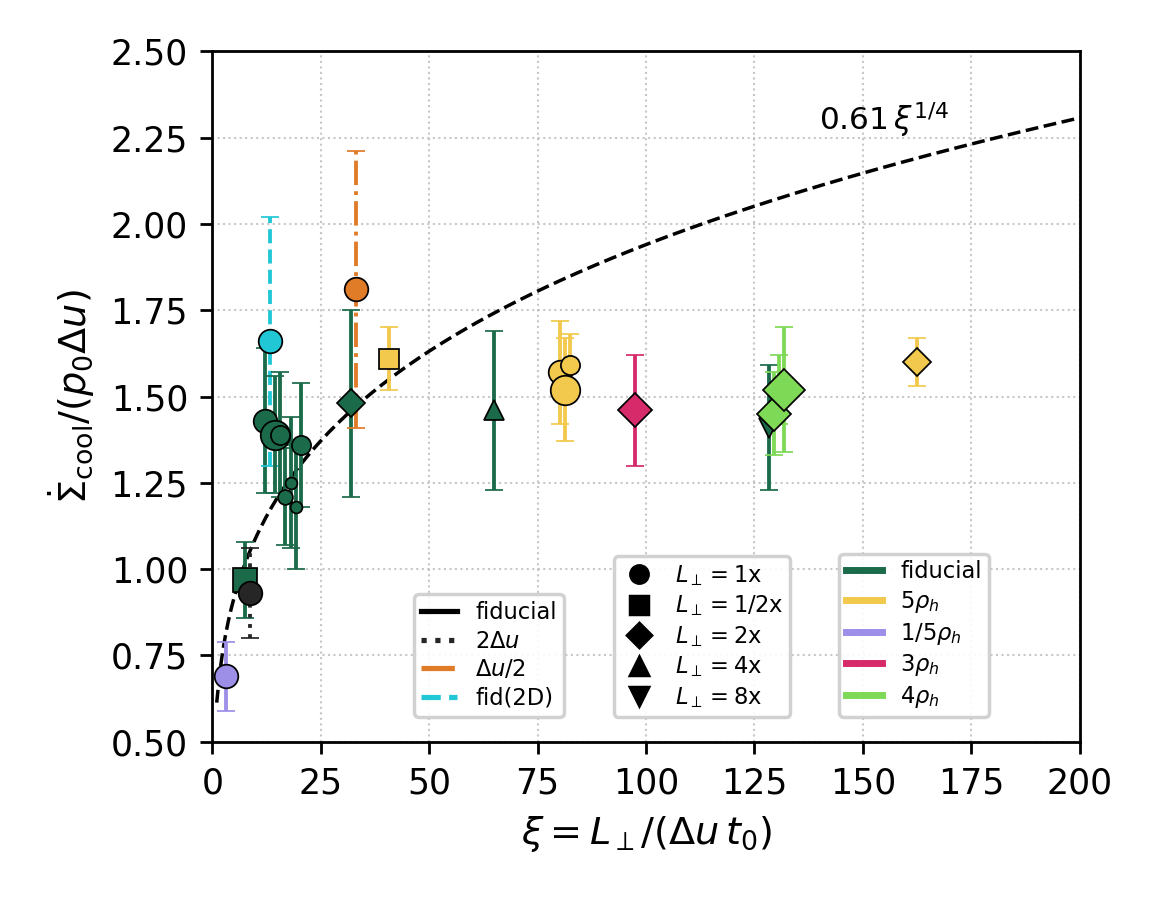}
\caption{\label{fig:sigma_cool_scaling}
Normalized cooling-rate surface density versus
$\xi=L_\perp/(\Delta u t_0)$, the shear to cooling times ratio.
Colors indicate $\rho_h$ (the $2\Delta u$ and $\Delta u/2$ runs are colored as their linestyles), marker shapes denote $L_\perp$, and linestyle labels
denote $\Delta u$; the cyan marker shows the fiducial 2D run, and larger markers indicate higher resolution.  
The dashed curve shows the $\xi^{1/4}$ scaling proposed in the literature (\cite{Fielding2020ApJ,Tan2021}), with the prefactor fitted to runs with $\xi \lesssim 17$. For $\xi \gtrsim 30$, $\dot{\Sigma}_{\rm cool}$ saturates well below it. Runs at $\xi \approx 81 $ and 130 agree across a factor of 4 in resolution.
}
\end{figure}

In stark contrast to earlier claims \cite{Fielding2020ApJ,Tan2021} that $\dot{\Sigma}_{\rm cool}/(p_0 \Delta u) \propto \xi^{1/4}$ \ps{($\xi = L_\perp/(\Delta u t_0)$ is the ratio of shear and cooling times)}, we find that for sufficiently large shear-to-cooling time ratio ($\xi \equiv L_\perp/[\Delta u t_0] \gtrsim 30$), the %transport coefficients and the 
entrainment or cooling rate density $\dot{\Sigma}_{\rm cool}$ and $z_0$ become independent of \ps{$\xi$ (Fig. \ref{fig:sigma_cool_scaling}); see section S5 \& Fig. 9 in SM}. This saturation has important implications for phenomenological models of multiphase galactic outflows \cite{Fielding2022ApJ,Dutta2025} and other radiative boundary layers where thermodynamics critically shapes momentum and energy transport \cite{Narasimha2011,Hillier2023}.

\emph{Data Availability}--- The authors agree to make any data required to support or replicate claims made in their article available privately to the journal’s editors, reviewers, and readers upon reasonable request. The problem generator, input, and other files used for our AthenaK simulations are publicly available on: \url{https://github.com/ArnavK04/Structure-of-Turbulent-Radiative-Mixing-Layers}.

\emph{Acknowledgements}--- This research was funded by the Anusandhan National Research Foundation (ANRF), Government of India, under the Advanced Research Grant (ARG) [Grant/Sanction Order No. ANRF/ARG/2025/002912/PS]. A.B. acknowledges the support of the Natural Sciences and Engineering Research Council of Canada (NSERC) through its Discovery Grant program. A.B. is grateful to the Department of Physics at the Indian Institute of Science for hosting his visit as Infosys Visiting Chair Professor. The simulations reported in this article were partly enabled by HPC resources provided by the Digital Research Alliance of Canada (alliancecan.ca) through an award to A.B. and by resources at Supercomputing Education and Research Center at the Indian Institute of Science. P.S. thanks useful discussions with Abhiram K., Jing Wang, and Xuchen Lin. Claude Opus 5.5 was used to edit text and generate figures. 

\bibliography{trml_refs}% Produces the bibliography via BibTeX.

\end{document}

\title{Supplemental Material for\\[4pt]
\textbf{``Structure of Turbulent Radiative Mixing Layers''}}

\maketitle

\section*{Overview}
This Supplemental Material provides additional details supporting the Letter
\textbf{``Structure of Turbulent Radiative Mixing Layers.''} Section~S1 introduces the physical parameters and the nondimensional Euler equations in the presence of radiative cooling. \ps{To contrast TRMLs from non-radiative mixing layers, Section~S2 presents a summary of the latter.}
Section~S3 describes our numerical setup and the list of simulations and their steady-state turbulent statistics. Section~S4 describes the structure of TRMLs and its parameter dependence. We also show additional diagnostics for selected runs. Section~S5 examines the scaling of the normalized radiative cooling-rate surface density,
\(\dot{\Sigma}_{\rm cool}/(p_0 \Delta u),\) with the dimensionless shear–cooling parameter
\(\xi = L_\perp/(\Delta u\, t_0).\)
In contrast to earlier works, our simulations show that in the strong–cooling
limit ($\xi \gg 1$) the surface brightness becomes independent of~$\xi$. \ps{Section~S6 presents results from a set of simulations that use different `engineered' cooling runs to assess generic features of TRMLs.} SM uses the same definitions as the Letter. 

%======================
\section{S1. Problem parameters \& nondimensional equations}
%======================

For two isobaric media with density contrast \(\chi = \rho_c/\rho_h = T_h/T_c\) and velocity shear \(\Delta u\), pressure balance implies \(\rho_c T_c = \rho_h T_h = \mu m_p p_0 / k_B\) \ps{($\mu m_p$ is the mean molecular mass and $k_B$ the Boltzmann constant)}. Using characteristic units \(\rho_h\) and \(\Delta u\), the Euler equations can be cast in terms of \(\chi\) and the Mach number \({\cal M} = \Delta u / c_{s,h}\), with \(c_{s,h} = (\gamma k_B T_h/\mu m_p)^{1/2}\).

We nondimensionalize time by the fiducial cooling time \(t_0\) and length by the shear–cooling scale \(\Delta u\,t_0\). The resulting equations of mass, momentum, and energy conservation \ps{in terms of dimensionless variables} are
\begin{subequations}
\begin{align}
    & \frac{\partial \tilde{\rho}}{\partial \tilde{t}} + \tilde{\nabla} \cdot (\tilde{\rho} \tilde{\mathbf{u}} ) = 0, \\
    & \frac{\partial}{\partial \tilde{t}} (\tilde{\rho} \tilde{\mathbf{u}}) 
    + \tilde{\nabla} \cdot \left( \tilde{\rho} \tilde{\mathbf{u}} \tilde{\mathbf{u}} 
    + \frac{\tilde{p}}{\gamma \mathcal{M}^2} \mathbf{I} \right) = 0, \\
    & \frac{\partial}{\partial \tilde{t}} \left( \tfrac{1}{2}\tilde{\rho} \tilde{u}^2 
    + \tfrac{\tilde{p}}{\gamma (\gamma - 1) \mathcal{M}^2} \right) 
    + \tilde{\nabla} \cdot \Big[ \Big(\tfrac{1}{2}\tilde{\rho} \tilde{u}^2
    + \tfrac{\tilde{p}}{(\gamma - 1)\mathcal{M}^2}\Big)\tilde{\mathbf{u}} \Big] = - \frac{\tilde{\rho}^2 \tilde{\Lambda}}{\gamma (\gamma - 1) \ps{\chi} \mathcal{M}^2},
\end{align}
\end{subequations}
where \(\tilde{p}=p/p_0\) and \(\tilde{\Lambda}=\Lambda(T)/\Lambda(T_0)\) \ps{($T_0 = \sqrt{T_h T_c}$ is the representative temperature of the mixing layer and $p_0$ is the pressure). The equations do not include explicit viscosity and conductivity but rely on numerical diffusion since microscopic dissipation is too small to resolve on computationally accessible grids.}

%======================
\section{S2. Structure of non-radiative mixing layers}
%======================

\begin{figure}
\includegraphics[scale=0.43]{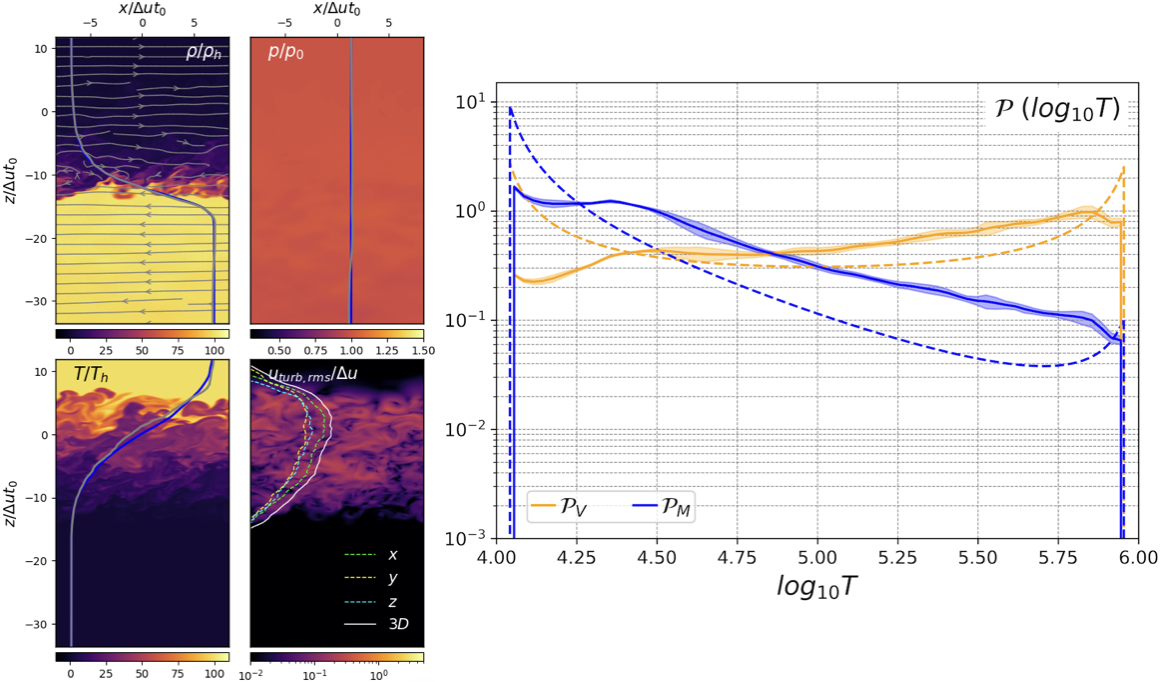} 
\caption{\label{fig:SML_fig1} Snapshots of density, pressure, temperature, and rms turbulent velocity (left panels) for the non-radiative shear mixing layer simulation ($t_0$ is taken $\approx 0.0194$ Myr, corresponding to our fiducial TRML run). The snapshots are at the same time as Fig. 1 in the Letter. Unlike TRMLs, velocity streamlines in the density panel show no net inflow from the diffuse to the dense phase. The hot and cold phases are increasingly mixed into a growing mixing layer. Pressure is almost constant throughout. The turbulent velocity has a broad distribution, unlike the TRML (Fig. 1 in the Letter) with a sharply peaked rms velocity. In TRMLs, turbulence spreads primarily into the dense phase, but without cooling, turbulence spreads into the diffuse phase. The blue (gray) solid lines show the instantaneous (time-averaged) vertical profiles. The right panel shows the volume- and mass-weighted PDFs for the same run. The solid lines are measured PDFs, and the dashed lines are based on a tanh fit (which fits poorly) and the isobaric assumption.}
\end{figure}

\begin{figure}
\includegraphics[scale=0.47]{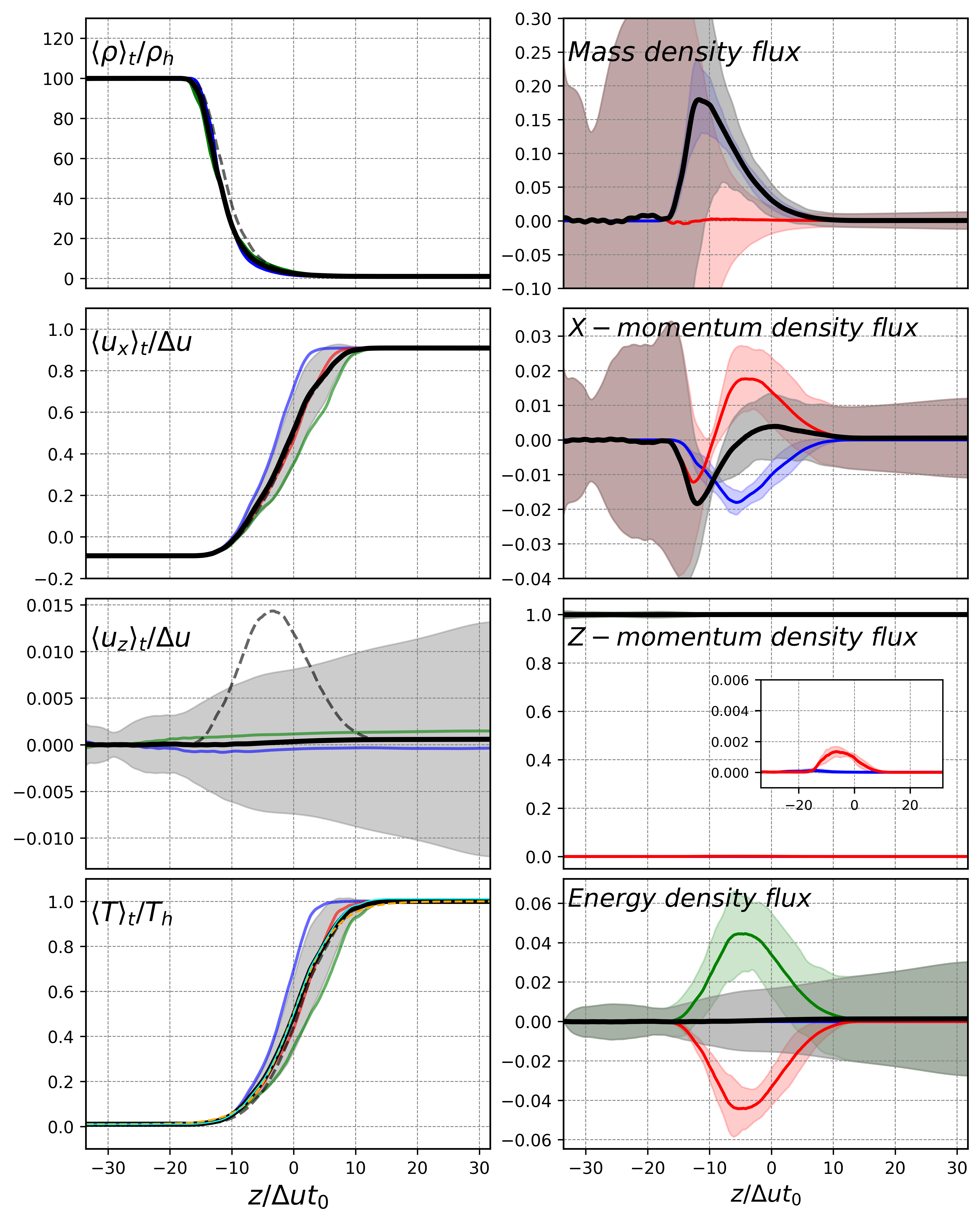}
\caption{\label{fig:SML_fig2} Mean vertical profiles of density, velocity components, temperature (left panels), and fluxes (right panels) for the non-radiative shear mixing layer simulation ($t_0 = 0.0194$ Myr). Shaded bands indicate 1$\sigma$ temporal variability. The profiles show similarity evolution and slowly progress into the hot/diffuse phase. The fluxes differ from TRMLs (the line styles and colors show the same quantities as in Fig. 2 of the Letter) and are consistent with the observed evolution of profiles. Turbulent fluxes are at least an order of magnitude smaller than in the TRMLs, with negligible ${\cal R}_{zz}$ flux relative to $p$ (inset in panel [3,2]). Unlike TRMLs, there is no net mass, momentum, and energy transport across the mixing layer.}
\end{figure}

To compare TRMLs with non-radiative shear mixing layers, we simulate the same physical parameters as our fiducial TRML run but without radiative cooling. Fig. \ref{fig:SML_fig1} shows snapshots at the same time as our fiducial TRML simulation (Fig. 1 in the Letter). The evolution without cooling is qualitatively different. The turbulent layer undergoes a similarity evolution with the width of the mixing layer increasing linearly with time, and primarily into the diffuse phase. The pressure dip in the mixing layer, which drives advection of the hot gas into the mixing layer in the presence of cooling, is absent without cooling. The peak velocity fluctuations are ten times smaller than with cooling.

The profiles and fluxes shown in Fig. \ref{fig:SML_fig2} also corroborate a qualitatively different evolution in shear mixing layers. Most notably, the net mass and energy fluxes vanish in the non-turbulent region, indicating neither mass nor energy transport across phases. The bumps in the net mass and x-momentum fluxes broaden the density and x-velocity profiles, respectively. Most notably, ${\cal R}_{zz}$ stress (and turbulent fluctuations in general) that was significant in TRMLs is negligible compared to the pressure.

%======================
\section{S3. Numerical setup and list of runs}
%======================

\begin{figure}
\includegraphics[scale=0.47]{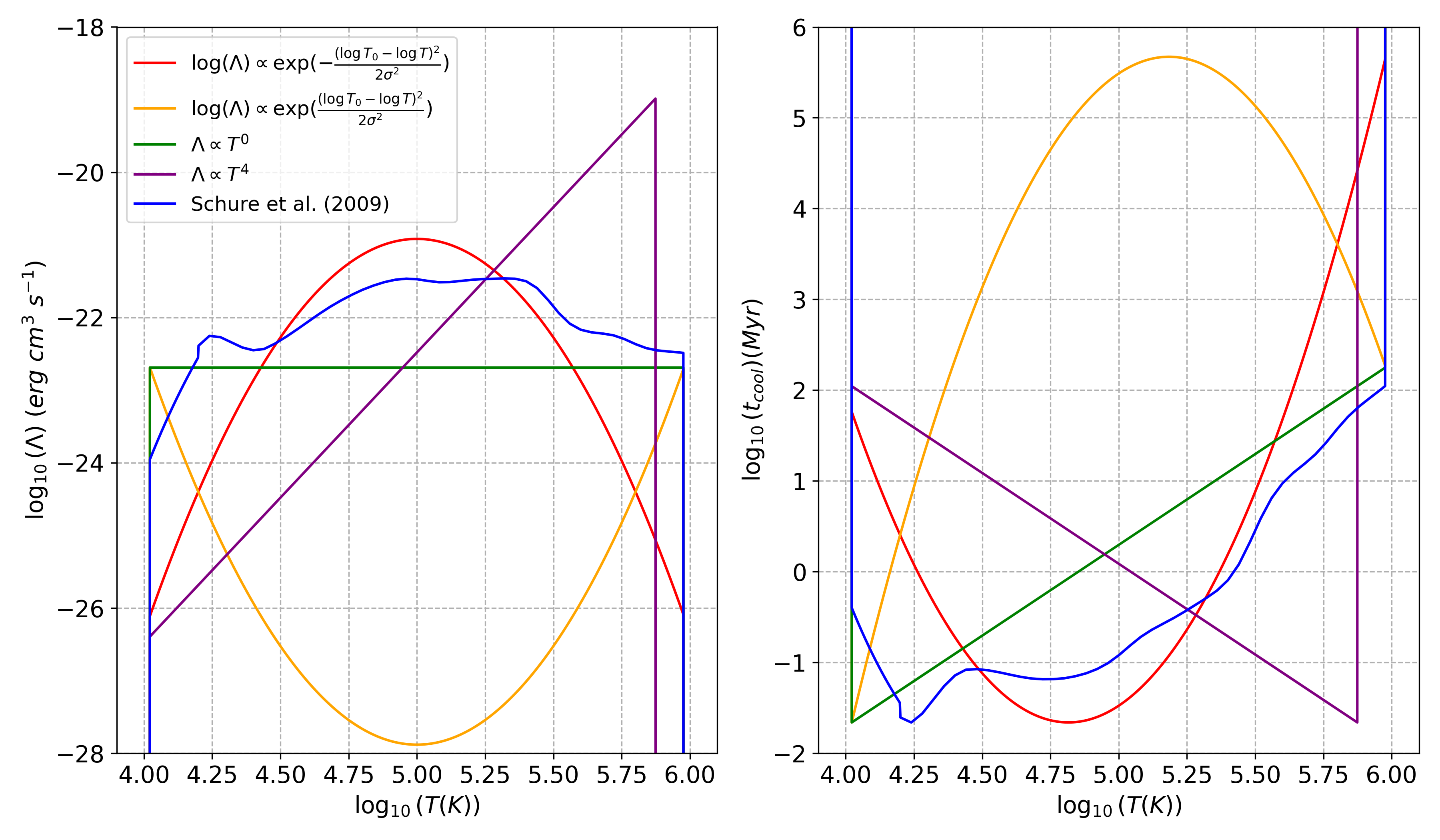}
\caption{\label{fig:cooling_functions}
The cooling functions $\Lambda(T)$ (left panel) and the isobaric cooling times ($t_{\rm cool} = 1.5 (k_B T)^2/(p \Lambda[T])$; right panel) as functions of temperature for our fiducial (\citet{Schure2009}) (the density/pressure here, corresponding to runs in Table \ref{tab:table_ecf}, is smaller than the fiducial run in Table \ref{tab:table}) and `engineered' cooling functions (see section S6). The minimum isobaric cooling time for all these cooling functions is the same, $\approx 0.022$ Myr, but the shapes differ dramatically.}
\end{figure}

Table~\ref{tab:table} lists simulations \ps{using our fiducial cooling function (blue curve in left panel of Fig. \ref{fig:cooling_functions})}, together with their measured steady–state properties. Initial and boundary conditions are described in the
Letter. We quote time–averaged quantities with their $1\sigma$ temporal
variations. The TRML thickness $z_0$ is obtained from a best–fit $\tanh$ profile
to the mean temperature (Eq.~(6) in the Letter). Across all simulations, the qualitative structure of
the TRML is similar; the tabulated peak turbulent correlations quantify the
strength of momentum and heat transport (see Fig.~2 of the Letter). Across all parameter variations, the turbulent stresses \({\cal R}_{zx}\) and \({\cal R}_{zz}\) retain their characteristic signs and orderings, reinforcing certain generality of the TRML structure.

\begin{table}[H]
\caption{\label{tab:table} List of runs and their turbulent statistics in steady state.}
\centering
\begin{threeparttable}
\begin{ruledtabular}
\scriptsize
\begin{tabular}{lcccccccc}
Run & $\xi$\tnote{c} & $u_{z,h}$ & $z_0$ & $\dot{\Sigma}_{\rm cool}$ & (max)~$\langle \delta \rho \delta u_z \rangle$\tnote{b} & (min)~${\cal R}_{zx}$\tnote{b} & (max)~${\cal R}_{zz}$\tnote{b} & (min)~${\cal Q}_t$\tnote{b} \\ 
 & & $(\Delta u)$ & $(\Delta u t_0)$ & $(p_0 \Delta u)$ 
 & $(\rho_h \Delta u)$ & $(\rho_h \Delta u^2)$ & $(\rho_h \Delta u^2)$ & $(p_0 \Delta u)$ \\ 
\hline
res-$280\times280\times1040$, fid\tnote{a} & 16.3 & $-0.57\pm 0.09$ & $2.17\pm 0.46$ & $1.43\pm 0.21$ & $12.93\pm 1.79$ & $-0.37\pm 0.12$ & $0.98\pm 0.24$ & $-0.29\pm 0.79$ \\
fid-2D & 16.3 & $-0.67 \pm 0.22$ & $3.75\pm 1.70$ & $1.66 \pm 0.36$ & $14.15 \pm 7.0$ & $-0.29\pm 0.74$ & $2.65 \pm 1.35$ & $-0.76 \pm 2.16$ \\
\hline
\multicolumn{9}{l}{\emph{[varying resolution]}} \\
res-2x & 16.3 & $-0.56\pm 0.08$ & $2.15 \pm 0.29$ & $1.39\pm 0.17$ & $12.76\pm1.99$ & $-0.33\pm 0.17$ & $1.09\pm 0.31$ & $-0.20\pm 0.66$ \\
res-1/2x & 16.3 & $-0.56\pm 0.07$ & $2.12\pm 0.43$ & $1.39\pm 0.18$ & $12.88\pm 2.58$ & $-0.32\pm 0.12$ & $0.95\pm 0.28$ & $-0.11\pm 0.58$ \\
res-$1/4$x & 16.3 & $-0.49\pm 0.06$ & $1.81\pm 0.40$ & $1.21\pm 0.14$ & $10.47\pm 1.75$ & $-0.29\pm 0.08$ & $0.73\pm 0.14$ & $-0.05\pm 0.11$ \\
res-$35\times35\times130$, res-$1/8$x & 16.3 & $-0.51 \pm 0.08$ & $2.23\pm 0.82$ & $1.25\pm 0.19$ & $11.53\pm 2.00$ & $-0.44\pm 0.16$ & $0.90\pm 0.17$ & $-0.09\pm 0.54$ \\
res-$32\times32\times128$, res $\approx$1/8x & 16.3 & $-0.48\pm 0.07$ & $1.84\pm 0.87$ & $1.18\pm 0.18$ & $10.97\pm 1.83$ & $-0.33\pm 0.09$ & $0.83\pm 0.12$ & $-0.07\pm 0.11$ \\
res$\approx1/16$x & 16.3 & $-0.40\pm 0.21$ & $0.84\pm 0.64$ & $0.97\pm 0.54$ & $7.95\pm 5.91$ & $-0.39\pm 0.31$ & $1.08\pm 0.71$ & $-0.07\pm 0.15$ \\
res$\approx1/32$x & 16.3 & $-0.45\pm 0.19$ & $0.78\pm 0.73$ & $1.09\pm 0.42$ & $1.61\pm 2.06$ & $-0.37\pm 0.29$ & $0.25\pm 0.37$ & $-0.01\pm 0.01$ \\
res$\approx1/64$x & 16.3 & $-0.80\pm 0.38$ & $3.09\pm 3.76$ & $1.93\pm 0.89$ & $7.39\pm 7.12$ & $-0.76\pm 0.44$ & $2.71\pm 0.93$ & $-0.09\pm 0.04$ \\
\hline
\multicolumn{9}{l}{\emph{[varying $L_\perp$]}} \\
$L_\perp$-1/2x & 8.1 & $-0.39\pm 0.06$ & $1.02\pm 0.16$ & $0.97\pm 0.11$ & $7.08\pm 1.50$ & $-0.24\pm 0.08$ & $0.50\pm 0.11$ & $-0.10\pm 0.41$ \\
res$\approx1/2$x, $L_\perp$-1x & 16.3 & $-0.55\pm0.08$ & $2.10\pm 0.47$ & $1.36\pm 0.18$ & $12.57\pm 2.15$ & $-0.32\pm 0.11$ & $0.86\pm 0.22$ & $-0.12\pm 0.46$ \\
res$\approx1/2$x, $L_\perp$-2x & 32.5 & $-0.60\pm 0.11$ & $3.23\pm 1.58$ & $1.48\pm 0.27$ & $14.76\pm 3.15$ & $-0.41\pm 0.11$ & $1.01\pm 0.28$ & $-0.63\pm 0.77$ \\
res$\approx1/2$x, $L_\perp$-4x & 65.0 & $-0.59 \pm 0.09$ & $3.21\pm 1.28$ & $1.46 \pm 0.23$ & $14.78 \pm 2.61$ & $-0.40 \pm 0.08$ & $1.03 \pm 0.24$ & $-0.51 \pm 0.46$ \\
res$\approx1/2$x, $L_\perp$-8x & 130 & $-0.57 \pm 0.07$ & $2.81\pm 0.9$ & $1.41 \pm 0.18$ & $13.72 \pm 1.9$ & $-0.38 \pm 0.06$ & $0.94 \pm 0.18$ & $-0.34 \pm 0.20$ \\
\hline
\multicolumn{9}{l}{\emph{[varying $\rho_h$]}} \\
$5\rho_h$ & 81.3 & $-0.64\pm 0.07$ & $3.51\pm 0.96$ & $1.57\pm 0.15$ & $15.90\pm 1.87$ & $-0.43\pm 0.03$ & $1.03\pm 0.11$ & $-0.10\pm 0.17$ \\
$5\rho_h$, res-$2$x & 81.3 & $-0.62\pm 0.08$ & $3.20\pm 0.93$ & $1.52\pm 0.15$ & $14.80\pm 1.95$ & $-0.41\pm 0.05$ & $0.96\pm 0.11$ & $-0.11\pm 0.14$ \\
$5\rho_h$, res-1/2x & 81.3 & $-0.65\pm 0.04$ & $3.18\pm 0.67$ & $1.59\pm 0.09$ & $15.70\pm 0.83$ & $-0.45\pm 0.04$ & $1.14\pm 0.05$ & $-0.01\pm 0.01$ \\
$5\rho_h$, res-1/2x, $L_\perp$-1/2x & 40.6 & $-0.66\pm 0.05$ & $3.11\pm 0.52$ & $1.61\pm 0.09$ & $15.84\pm 1.41$ & $-0.48\pm 0.10$ & $1.10\pm 0.10$ & $-0.02\pm 0.01$ \\
$5\rho_h$, res-1/2x, $L_\perp$-2x & 163 & $-0.66\pm 0.04$ & $3.31\pm 0.76$ & $1.60\pm 0.07$ & $15.96\pm 1.05$ & $-0.45\pm 0.04$ & $1.13\pm 0.06$ & $-0.01\pm 0.004$ \\
$\rho_h/5$ & 3.3 & $-0.28\pm 0.04$ & $0.56\pm 0.11$ & $0.69\pm 0.10$ & $3.90\pm 1.20$ & $-0.15\pm 0.04$ & $0.29\pm 0.07$ & $-0.095\pm 0.25$ \\
$4\rho_h,\ L_\perp$-2x & 130 & $-0.59\pm 0.05$ & $2.97\pm 0.74 $ & $1.45\pm 0.12$ & $14.37\pm 1.29$ & $-0.37\pm 0.03$ & $0.90\pm 0.07$ & $-0.05\pm 0.14$ \\
$4\rho_h$, res-1/2x, $L_\perp$-2x & 130 & $-0.62\pm 0.04$ & $3.17\pm 0.85$ & $1.52\pm 0.10$ & $14.56\pm 1.64$ & $-0.40\pm 0.02$ & $0.97\pm 0.08$ & $-0.03\pm 0.06$ \\
$4\rho_h$, res-2x, $L_\perp$-2x & 130 & $-0.62\pm 0.07$ & $3.25\pm 1.05 $ & $1.52\pm 0.18$ & $14.41\pm 2.12$ & $-0.39\pm 0.05$ & $0.94\pm 0.18$ & $-0.26\pm 0.21$ \\
$3\rho_h,\ L_\perp$-2x & 97.6 & $-0.59\pm 0.07$ & $3.06\pm 0.87 $ & $1.46\pm 0.16$ & $14.91\pm 2.11$ & $-0.39\pm 0.04$ & $0.94\pm 0.15$ & $-0.19\pm 0.30$ \\
\hline
\multicolumn{9}{l}{\emph{[varying $\Delta u$]}} \\
$2\Delta u$ & 8.1 & $-0.37\pm 0.06$ & $1.06\pm 0.29$ & $0.93\pm 0.13$ & $7.09\pm 0.90$ & $-0.21\pm 0.05$ & $0.42\pm 0.07$ & $-0.19\pm 0.45$ \\
$\Delta u/2$ & 32.5 & $-0.73\pm 0.17$ & $3.98\pm 1.19$ & $1.81\pm 0.40$ & $17.12\pm 5.11$ & $-0.44\pm 0.18$ & $1.56\pm 0.55$ & $-0.23\pm 1.21$ \\
\end{tabular}
\normalsize
\end{ruledtabular}
\begin{tablenotes}
\item[a]{Fiducial run with $\rho_h/m_p = 10^{-3}\,\mathrm{cm}^{-3}$, $\Delta u = 31\,\mathrm{km\,s^{-1}}$, $\chi = 100$, resolution $280\times280\times1040$, and box size $[10,10,40]$ pc. The cooling time $t_0$ for our fiducial run is 0.0194 Myr. Other runs vary resolution, transverse box-size ($L_\perp$), $\rho_h$, and/or $\Delta u$. \ps{All our runs are evolved for $\approx 500 t_0$.}}
\item[b]{Maximum/minimum values of the relevant turbulent correlations measured within the TRML; the shapes of these profiles are similar across all runs (see Fig.~2 of the Letter).}
\item[c]{Shear-to-cooling time ratio $\xi = L_\perp/(\Delta u\, t_0)$; $\xi \approx 16.3$ for the fiducial run and scales as $L_\perp \rho_h/\Delta u$ (since $t_0 \propto \rho_h^{-1}$ at fixed temperatures).}
\end{tablenotes}
\end{threeparttable}
\end{table}

%======================
\section{S4. Structure of TRMLs \& parameter dependence}
%======================

\begin{figure}%[b]
\includegraphics[scale=0.5]{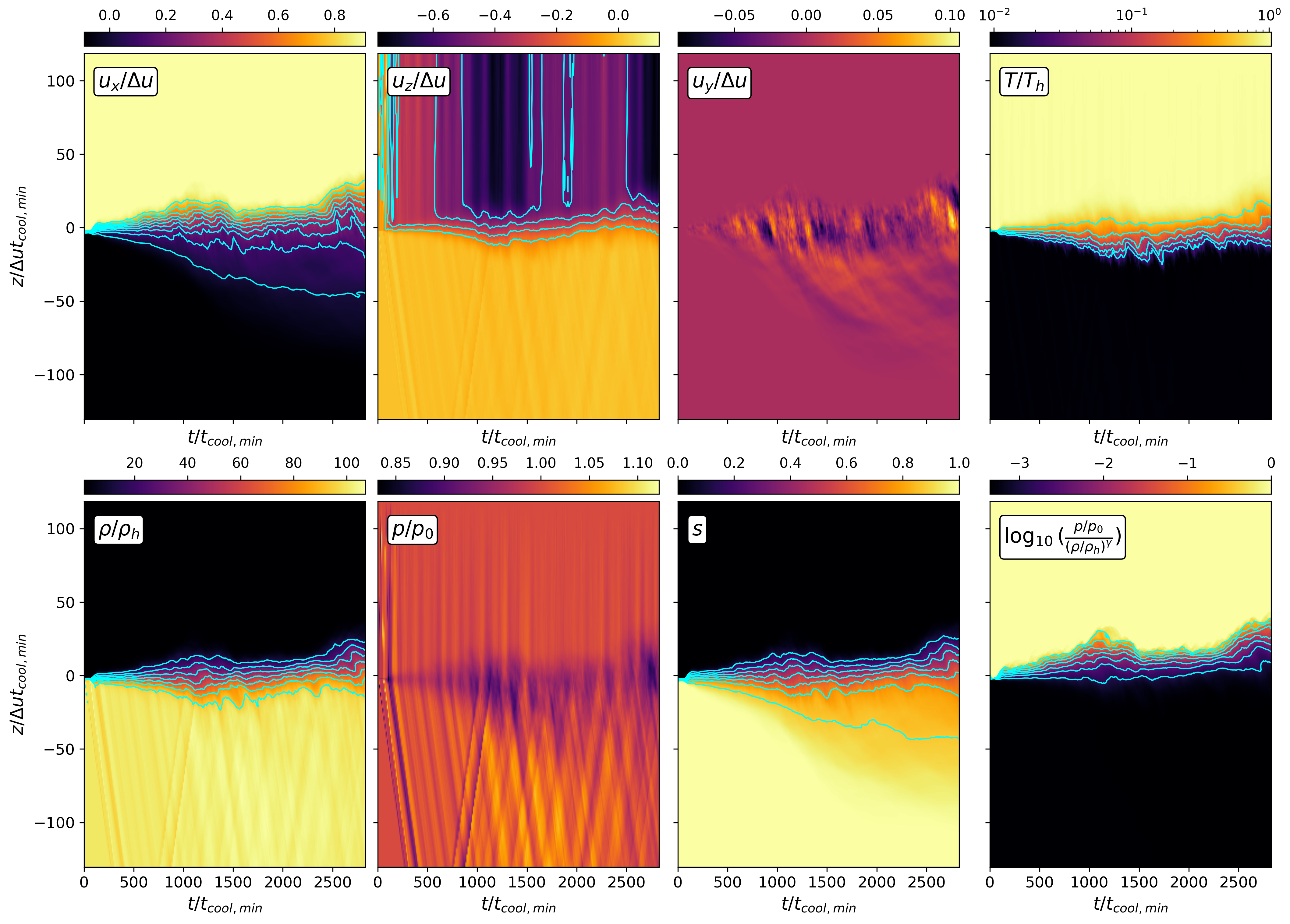}
\caption{\label{fig:xt} \ps{Space-time diagrams of various horizontally averaged quantities (shown in color; cyan lines show contours for clarity) as a function of $z$ and time in the simulation frame for the fiducial run. Prominent streaks in the pressure panel correspond to sound waves. Also note the pressure dip in the TRML and the steady vertical advection velocity in the hot phase. The simulation domain is taller than the vertical extent shown. Time is normalized with $t_{\rm cool, min}$ instead of $t_0$ ($t_{\rm cool, min} = t_0/5.49$).}}
\end{figure}

\ps{Fig. \ref{fig:xt} shows a space-time diagram for our fiducial TRML run that stacks the mean profiles over time. As shown in the Letter (Fig. 2, $\langle u_x \rangle$ panel), the shear layer splits into two parts -- the steady thermal layer (stationary in TRML frame and slowly rising up in the simulation frame) and the $x-$ momentum layer that slowly spreads into the dense phase. The average horizontal velocity $\langle u_x \rangle$ does not reach a steady state because, over time, momentum is added to the TRML. A similar spreading is seen for the space-time diagram of the passive scalar ($s$) because, like $u_x$ (Eq. 3 in Letter), it also satisfies 
\[
\frac{d}{dt} \int \langle s \rho \rangle dz = -\langle \rho u_z \rangle \Delta s,
\]
where $\Delta s = s_h - s_c$, the passive scalar jump across the TRML. Notice the sound waves in the pressure snapshot, which also shows a clear pressure dip in the mid-plane responsible for entrainment.}

\begin{figure}
\includegraphics[scale=0.47]{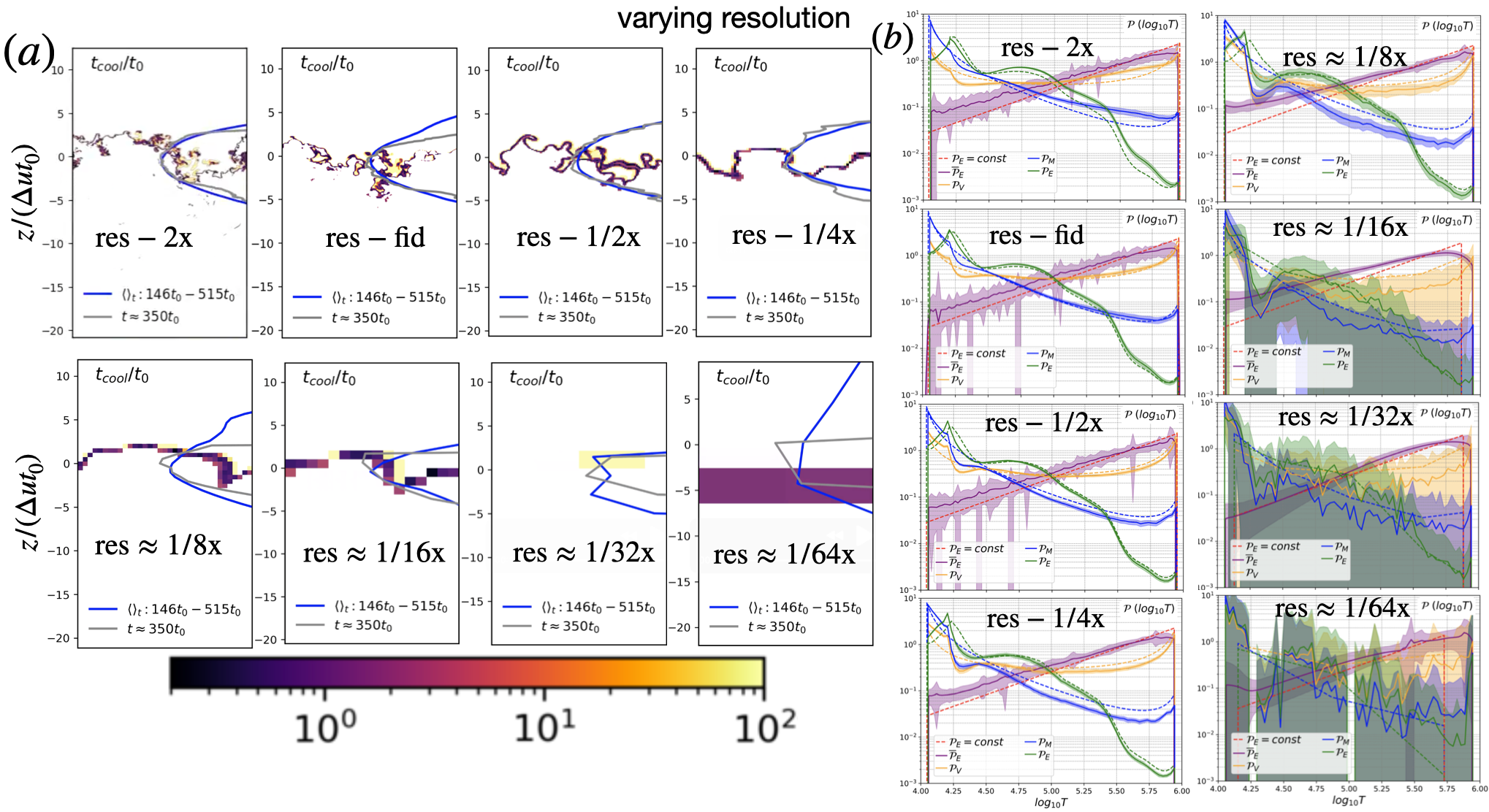}
\caption{\label{fig:vary_resolution}
(a) Cooling-time snapshots at different resolutions. Lower resolution produces less wrinkled interfaces, as expected from Eq. \ref{eq:delta_by_z0}. 
(b) Volume, mass, and emissivity PDFs (see Fig.~3 of the Letter). Significant
temporal variability appears only once the TRML becomes under-resolved ($\lesssim 1/8$ of the fiducial resolution), but even then, the time-averaged PDFs remain broadly consistent.}
\end{figure}

\begin{figure}
\includegraphics[scale=0.8]{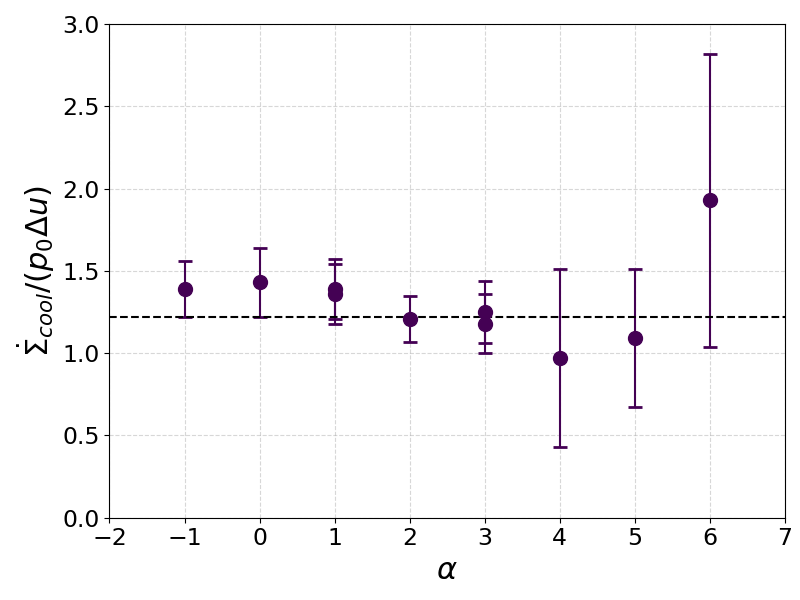}
\caption{\label{fig:dot_Sigma_vs_res}
Normalized cooling-rate surface density $\dot{\Sigma}_{\rm cool}$ versus grid
resolution parameter $\alpha$ (where grid resolution $\Delta = 2^\alpha\Delta_{\rm fid}$). The three highest resolutions agree very well. Even the poorest-resolution run
($1/64$ of fiducial) yields a time-averaged surface brightness consistent within
$1\sigma$, although with large temporal fluctuations.}
\end{figure}

\begin{figure}
\includegraphics[scale=0.35]{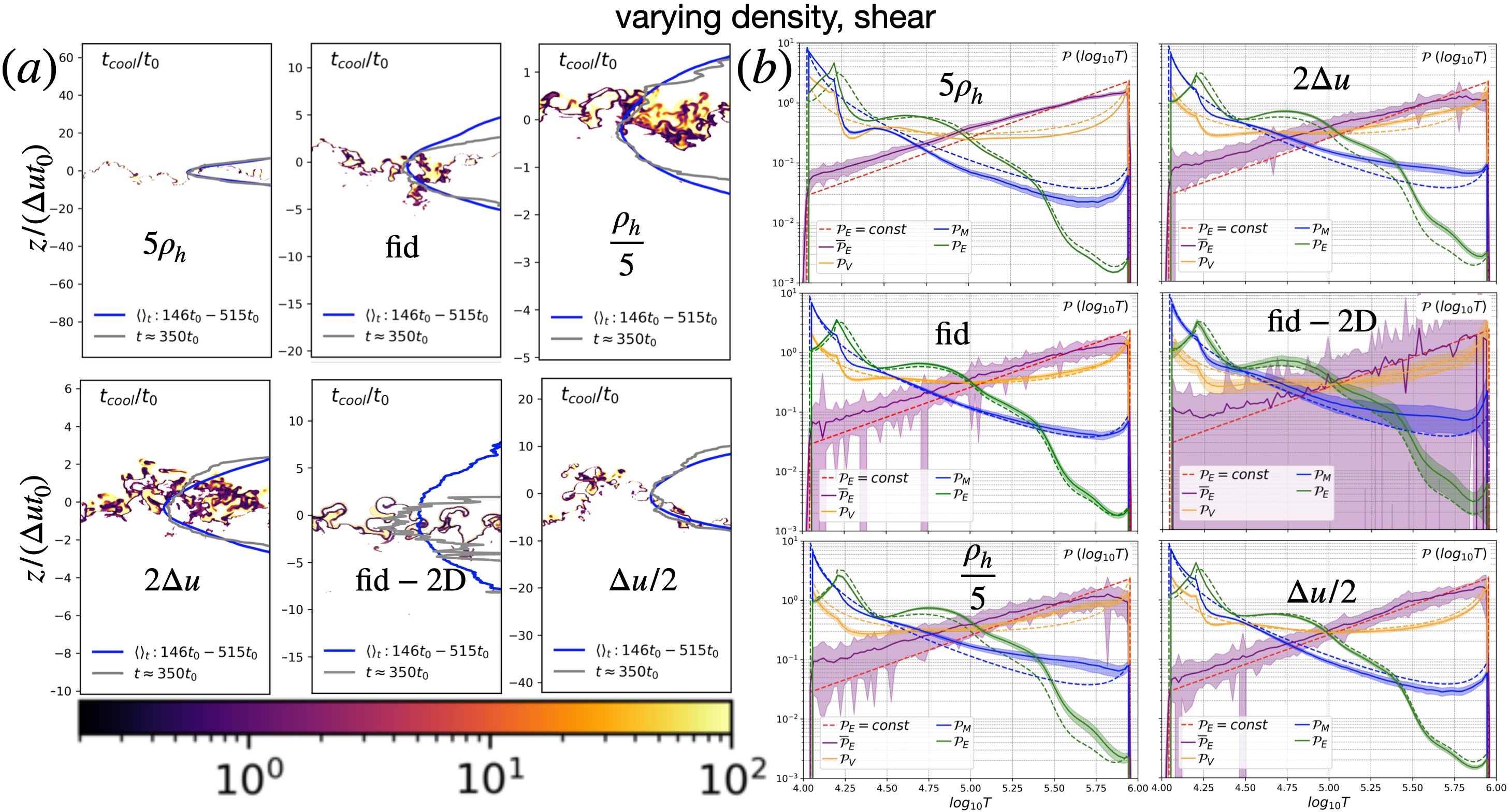}
\caption{\label{fig:vary_rho_delu}
(a) Cooling-time snapshots for different $\rho_h$ and $\Delta u$ and for the
fiducial 2D run. Smaller $\rho_h$ or larger $\Delta u$ (i.e.\ smaller
$\xi=L_\perp/[\Delta u t_0]$) produces more warm gas and a more contorted cooling
interface.  
(b) Corresponding PDFs. The 2D run shows unusually large temporal variability,
consistent with the corresponding panel in (a).}
\end{figure}

\begin{figure}
\includegraphics[scale=0.34]{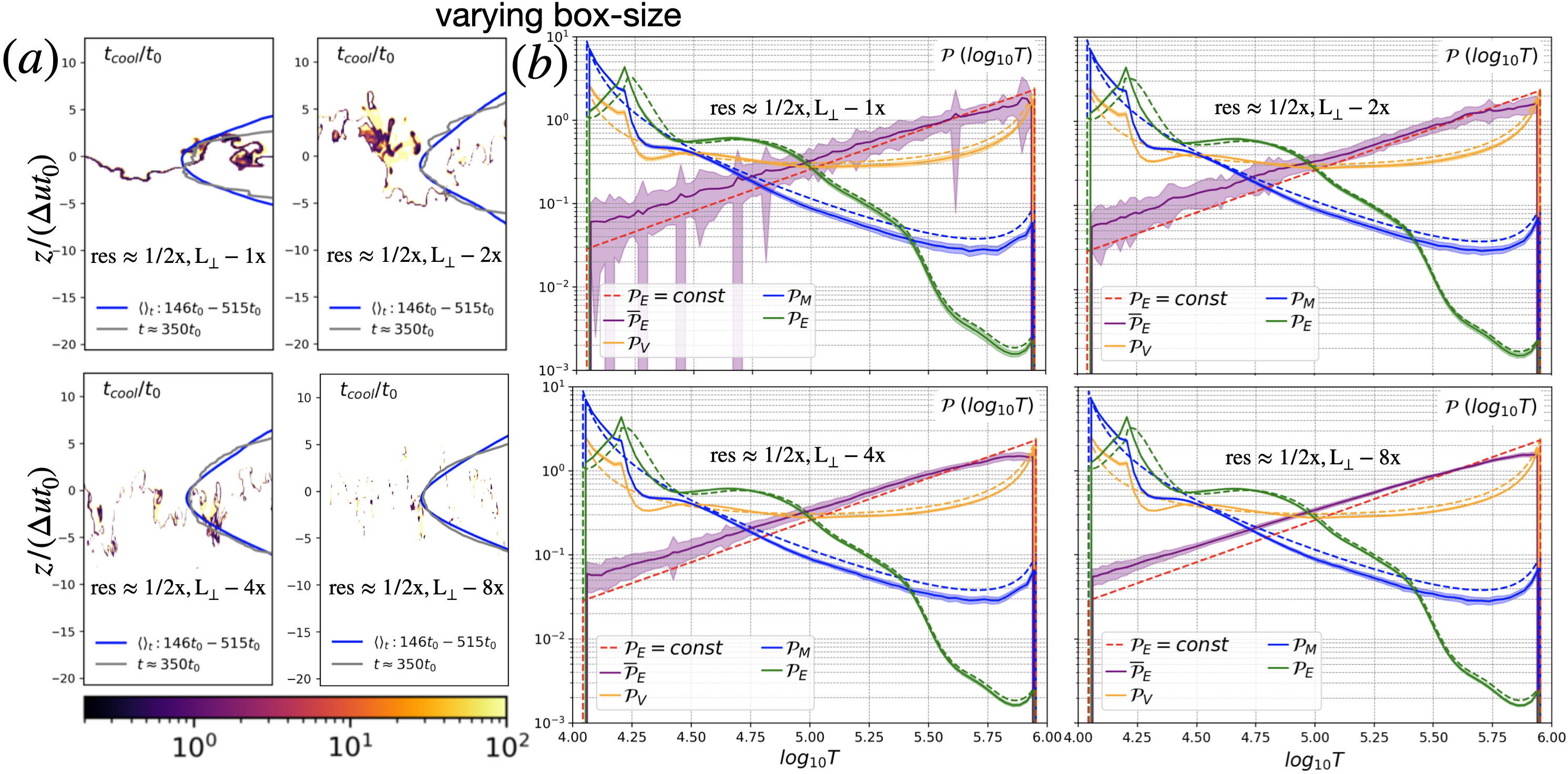}
\caption{\label{fig:vary_Lperp}
(a) Cooling-time snapshots for varying $L_\perp$. For small $L_\perp$, the
dominant wavelength of the cooling interface is of order $L_\perp$; for larger
$L_\perp$, the eddy scale saturates while $L_\perp$ increases, so $z_0 \ll
L_\perp$.  
(b) Corresponding PDFs. Temporal fluctuations diminish with increasing
$L_\perp$ due to averaging over a larger number of grid cells.}
\end{figure}

The cooling layer, which has a fractal structure \cite{Fielding2020ApJ}, becomes thinner and more geometrically complex at higher resolution. However, the net cooling rate is converged within $1\sigma$ even for extremely coarse grids (Fig.~\ref{fig:dot_Sigma_vs_res}). As shown in Fig. 1 of the Letter and Figs.~\ref{fig:vary_resolution}a,~\ref{fig:vary_rho_delu}a, and~\ref{fig:vary_Lperp}a, radiative losses concentrate in corrugated surfaces far thinner than the TRML itself ($z_0 \simeq 2.17 \Delta u t_0$ in the fiducial run). The total cooling rate is $\approx \langle n^2 \Lambda \rangle A_g z_0$ ($A_g = L_\perp^2$ is the geometric cross-sectional area of the domain), which can be recast as $(n^2 \Lambda)_p A_F \delta$ with $(n^2 \Lambda)_p$ denoting peak cooling-rate density, $A_F$ the fractal surface area, and $\delta$ the interface thickness. Approximating the integrated cooling rate as $\dot{\Sigma}_{\rm cool} \simeq \langle n^2 \Lambda \rangle z_0$ gives
\begin{equation}
\label{eq:delta_by_z0}
\frac{\delta}{z_0} \approx \frac{\dot{\Sigma}_{\rm cool} k_B^2}{z_0 p^2 (\Lambda/T^2)_p} \, \frac{A_g}{A_F},
\end{equation}
which for our fiducial run yields $\delta/z_0 \simeq 0.09\, A_g/A_F \ll 1$. Because $A_F/A_g$ grows with numerical resolution \cite{Fielding2020ApJ}, for the volume of rapidly cooling gas and $\dot{\Sigma}_{\rm cool}$ to remain resolution-independent, the fractal area increase must be compensated by the decrease in $\delta$, and $z_0$ must be resolved (while $\delta$ may not be).\footnote{The TRML thickness is $z_0\simeq 2.17\,\Delta u t_0$ and is resolved by $\gtrsim 4$ cells for all runs with resolution $\gtrsim 1/8$ of fiducial.} TRML thickness $z_0$ is larger than the much-thinner fractal cooling layers or laminar cooling flows, which require $\delta$ to be resolved \cite{Dutta2022}. Moreover, since the fractal area is much larger than the geometric area, the TRML surface brightness is expected to exceed that of a uniform cooling flow by $\approx A_F/A_g$.

Figs.~\ref{fig:vary_resolution},~\ref{fig:vary_rho_delu},
and~\ref{fig:vary_Lperp} show cooling–time snapshots and corresponding PDFs
for variations in resolution, density, shear, and transverse box size $L_\perp$.
The PDFs exhibit convergence across all well-resolved runs. As
resolution improves, the warm intermediate-temperature gas becomes more prominent because smaller eddies mix faster than the shortest cooling times \cite{Lancaster2026}. Nevertheless, for the fiducial parameters, the variation in $\dot{\Sigma}_{\rm cool}$ (Fig. \ref{fig:dot_Sigma_vs_res}) and mean PDFs (Fig. \ref{fig:vary_resolution}) across more than two orders of magnitude in resolution is modest. The variability in both $\dot{\Sigma}_{\rm cool}$ and the PDFs increases dramatically at the three lowest resolutions, where $z_0$ is poorly resolved.

%======================
\section{S5. Scaling of $\dot{\Sigma}_{\rm cool}$ with $\xi$}
%======================

A key result of this work is that the normalized cooling-rate surface density $\dot{\Sigma}_{\rm cool}/(p_0\Delta u)$ saturates relative to $\xi$ ($=L_\perp/(\Delta u t_0)$, the ratio of the shearing time to the cooling time) in the strong-cooling regime ($\xi \gg 1$). Earlier studies \cite{Fielding2020ApJ, Tan2021} argued for a scaling $\dot{\Sigma}_{\rm cool} \propto \xi^{1/4}$, but our results (Fig.~5 in the Letter) show no such dependence for $\xi\gg1$. Indeed, a similar flattening at large \(\xi\) is visible in Fig. 8 of \citet{Tan2021}, although it was not emphasized there.

\begin{figure}
\includegraphics[scale=0.75]{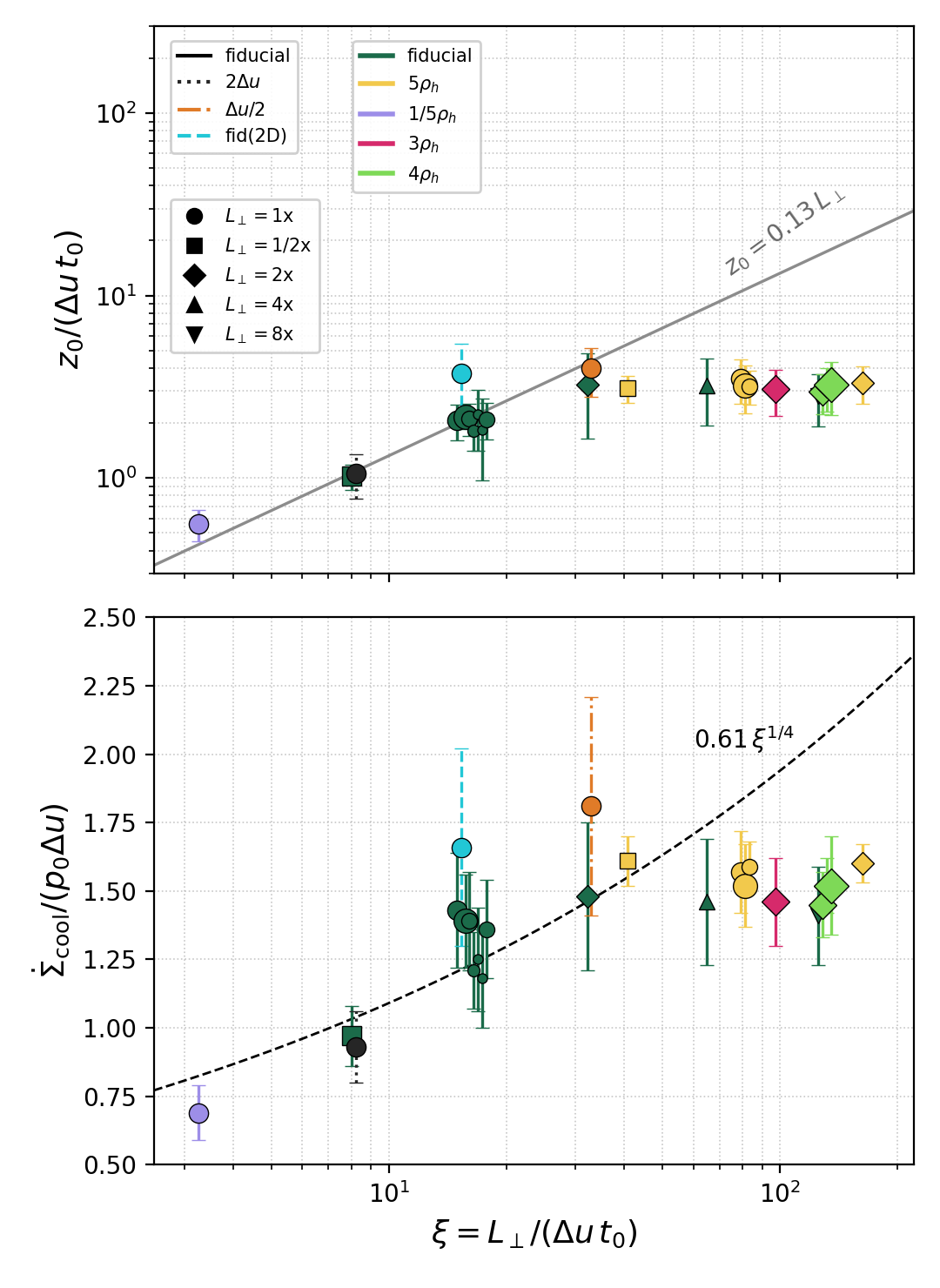}
\caption{\label{fig:z_0_sigma_cool_scaling} TRML thickness $z_0/(\Delta u\, t_0)$ (top) and normalized cooling-rate
surface density $\dot{\Sigma}_{\rm cool}/(p_0 \Delta u)$ (bottom) versus the
shear-to-cooling time ratio $\xi = L_\perp/(\Delta u\, t_0)$ for all runs in
Table~\ref{tab:table} with resolution $\geq 1/8$ of fiducial. Symbols follow
Fig.~5 of the Letter: colors denote $\rho_h$ (the $2\Delta u$ and $\Delta u/2$
runs are colored as their linestyles), marker shapes denote $L_\perp$, error-bar
linestyles denote $\Delta u$, the cyan marker shows the fiducial 2D run, and
larger markers indicate higher resolution. Error bars show $1\sigma$ temporal
variability; runs with nearly equal $\xi$ are slightly offset horizontally for
clarity. The grey line ($z_0 = 0.13\, L_\perp$) and the dashed curve
($0.61\,\xi^{1/4}$) are weighted least-squares fits to the 3D runs with
$\xi \lesssim 17$. Runs reaching the same $\xi$ through different parameters
($L_\perp$-1/2x and $2\Delta u$ at $\xi \approx 8$; $L_\perp$-2x and $\Delta u/2$
at $\xi \approx 33$; $L_\perp$-8x and $4\rho_h$ at $\xi \approx 130$) agree within
errors, indicating that $\xi$ controls the TRML structure. At small $\xi$, the layer
thickness is limited by the box size ($z_0 \propto L_\perp$) and
$\dot{\Sigma}_{\rm cool}$ follows $\xi^{1/4}$. For $\xi \gtrsim 30$, both $z_0$ and
$\dot{\Sigma}_{\rm cool}$ saturate, at $z_0 \approx 3\,\Delta u\, t_0$ and
$\dot{\Sigma}_{\rm cool} \approx 1.5\, p_0 \Delta u$, while $L_\perp/z_0$ grows to
$\approx 50$ at the largest $\xi$. Turbulent transport is then set by $z_0$ rather
than $L_\perp$.
}
\end{figure}

If instead $\xi$ is defined using the inertial scale of turbulence rather than the box width---equivalent to the Damk\"ohler number in combustion---the $\xi^{1/4}$ scaling seems to reappear \cite{Lancaster2026}. While physically appealing and tied to the small-scale structure of TRMLs, this definition requires measuring the turbulent velocity power spectrum, whereas our parameter $\xi=L_\perp/(\Delta u t_0)$ is a global dimensionless number, readily calculated from the setup parameters.

The physical origin of the saturation is intuitive. Unlike non-radiative shear layers, which grow indefinitely and become box-size limited, TRMLs in the strong-cooling regime develop a vertically thin layer of thickness $z_0\ll L_\perp$ whose dominant eddy scale is also $\sim z_0$. As Fig.~\ref{fig:vary_Lperp} shows, the eddies governing transport remain far smaller than $L_\perp$ even as the domain widens. Turbulent transport, therefore, depends only on $z_0$ and not on $L_\perp$, yielding the observed $\xi$-independent surface brightness. By varying the gas density, shear, and box-size, we verify this scaling (Figs. \ref{fig:vary_rho_delu}, \ref{fig:vary_Lperp}). Fig. \ref{fig:z_0_sigma_cool_scaling} shows the scaling of the TRML thickness ($z_0$; top panel) and $\dot{\Sigma}_{\rm cool}$ as a function of $\xi$. Both $z_0$ and $\dot{\Sigma}_{\rm cool}$ become independent of $\xi$ for $\xi \gtrsim 30$, consistent with the saturation of turbulent eddies in TRMLs at scales comparable to $z_0$, and much smaller than the box-size.

%======================
\section{S6. `Engineered' cooling functions}
%======================
\label{sec:engineered_CF}

\begin{table}[H]
\caption{\label{tab:table_ecf} List of runs using `engineered' cooling functions with $t_{\rm cool, min} = 0.022$ Myr and their turbulent statistics.}
\centering
\begin{threeparttable}
\begin{ruledtabular}
\scriptsize
\tiny
\begin{tabular}{lcccccccc}
% & & & &  max\tnote{b}  & min\tnote{b} & max\tnote{b} & min\tnote{b} \\ 
Cooling function & $t_0/t_{\rm cool,min}$\tnote{a} & $u_{z,h}$ & $z_0$ & $\dot{\Sigma}_{\rm cool}$ & (max)~$\langle \delta \rho \delta u_z \rangle$ & (min)~${\cal R}_{zx}$ & (max)~${\cal R}_{zz}$ & (min)~${\cal Q}_t$ \\
&  & $(\Delta u)$ & $(\Delta u t_{\rm cool,min})$ & $(p_0 \Delta u)$ & $(\rho_h \Delta u)$ & $(\rho_h \Delta u^2)$ & $(\rho_h \Delta u^2)$ & $(p_0 \Delta u)$ \\ 
\hline
\multicolumn{9}{l}{\emph{[res-$256\times256\times1024$, ($f_c=1/3 $)]}} \\
Schure et al. & $5.49$ &$-0.234\pm 0.054$ & $2.31\pm 0.45$& $0.590\pm 0.081$ & $4.98\pm 0.95$ & $-0.171\pm 0.048$ & $0.330\pm 0.064$ & $-0.26\pm 0.24$ \\
Schure et al., res-1/2x& $5.49$ &$-0.242\pm 0.047$ & $2.22\pm 0.38$& $0.604\pm 0.086$ & $4.82\pm 0.88$ & $-0.168\pm 0.053$ & $0.323\pm 0.074$ & $-0.22\pm 0.32$ \\
Log-normal & $1.53$ &$-0.291\pm 0.063$ & $1.93\pm 0.36$& $0.731\pm 0.088$ & $3.68\pm 0.88$ & $-0.187\pm 0.037$ & $0.379\pm 0.086$ & $-0.25\pm 0.23$ \\
Log-normal, res-1/2x&  $1.53$ &$-0.276\pm 0.051$ & $1.77\pm 0.29$& $0.695\pm 0.065$ & $3.39\pm 0.45$ & $-0.172\pm 0.039$ & $0.331\pm 0.046$ & $-0.19\pm 0.23$ \\
Inv Log-normal & $1.41\times 10^7$& $-0.047 \pm 0.015$&$3.08 \pm 0.43$ &$0.102 \pm 0.012$ & $1.35\pm 0.15$&$-0.047\pm 0.008$ &$0.075\pm 0.009$ &$-0.12\pm 0.03$ \\
Inv Log-normal, res-1/2x &  $1.41\times 10^7$ & $-0.049 \pm 0.017$&$ 3.97\pm0.43$ &$0.098 \pm 0.007$ & $1.43\pm 0.21$&$-0.040\pm 0.008$ &$0.071\pm 0.008$ &$-0.10\pm 0.05$ \\
Constant & $90.36$ &$-0.133\pm 0.040$ & $2.58\pm 0.48$& $0.322\pm 0.046$ & $3.16\pm 0.41$ & $-0.099\pm 0.027$ & $0.211\pm 0.045$ & $-0.21\pm 0.18$ \\
Constant, res-1/2x&  $90.36$ &$-0.179\pm 0.046$ & $2.89\pm 0.29$& $0.427\pm 0.046$ & $4.32\pm 0.69$ & $-0.118\pm 0.055$ & $0.281\pm 0.077$ & $-0.17\pm 0.18$ \\
Hotpeak& $55.94$ &$-0.148\pm 0.027$ & $1.82\pm 0.24$& $0.369\pm 0.027$ & $0.34\pm 0.06$ & $-0.097\pm 0.021$ & $0.128\pm 0.011$ & $-0.09\pm 0.04$ \\
Hotpeak, res-1/2x& $55.94$ &$-0.175\pm 0.025$ & $1.94\pm 0.27$& $0.435\pm 0.035$ & $0.39\pm 0.06$ & $-0.091\pm 0.016$ & $0.136\pm 0.013$ & $-0.10\pm 0.07$ \\
\end{tabular}
\normalsize
\end{ruledtabular}
\begin{tablenotes}
    \item[a] These runs have the same minimum isobaric cooling time across temperatures (see Fig. \ref{fig:cooling_functions}) but widely varying $t_0$ (=0.12 Myr for fiducial $\Lambda$ in this table). The box-size, shear velocity, and density contrast are the same as our fiducial run in Table \ref{tab:table}. Unlike Table \ref{tab:table} where the initial cold gas volume  fraction was $f_c=2/3$, the initial cold gas fraction here is 1/3 (since the gas is less radiative and $\xi \approx 2.6$). Turbulent mass transport is much lower with Hotpeak $\Lambda$ that peaks at higher temperatures (lower densities).
\end{tablenotes}
\end{threeparttable}
\end{table}

Primarily to test the remarkable coincidence for our fiducial cooling function, namely that the volume PDF across horizontally-averaged temperature equals the PDF across the physical temperature (see Fig. 3 in the Letter), we run a set of `engineered' cooling functions shown in Fig. \ref{fig:cooling_functions} and Table \ref{tab:table_ecf}. These cooling functions ($\Lambda$s) differ substantially. The lognormal $\Lambda(T)$, like the fiducial  $\Lambda$, has the shortest cooling time for the temperatures between $T_c$ and $T_h$. The inverse-lognormal $\Lambda$ has long cooling times at intermediate temperatures. Constant ($\Lambda \propto T^0$) and hot-peaked ($\Lambda \propto T^4$) cooling functions have the shortest cooling time at $T_c$ and $T_h$, respectively.

\begin{figure}
\includegraphics[scale=0.34]{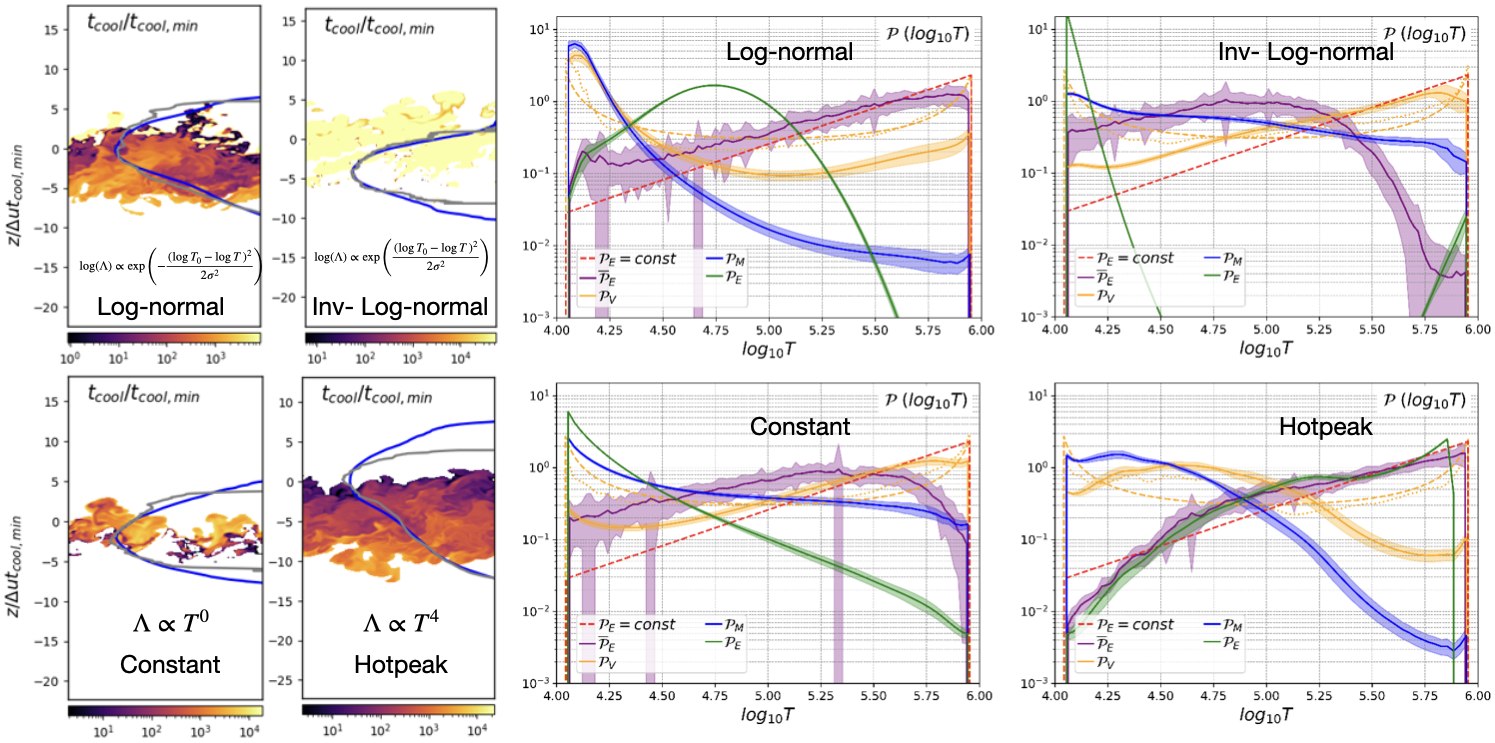}
\caption{\label{fig:eng_cool} Steady-state snapshots of cooling times (left panels; instead of $t_0$ we scale the axes with $t_{\rm cool, min} \approx 0.022$ Myr) and the volume, mass, and emissivity PDFs versus the horizontally-averaged ($\langle T \rangle$) and physical temperatures ($T$) for the simulations with engineered cooling functions. Compare the cooling time snapshot with the bottom-left panel of Fig. 1 in the Letter for the fiducial $\Lambda$. The solid lines in the PDF panels show the volume (orange), mass (blue), and emissivity (green) PDFs across physical temperature ($T$). The dotted orange line is the volume PDF as a function of the horizontally-averaged temperature $\langle T \rangle$, and the dashed orange line is based on a best-fit tanh profile to $\langle T \rangle$. Clearly, the volume PDFs across $T$ and $\langle T \rangle$ differ substantially. The magenta line shows the emissivity PDF across $\langle T \rangle$, which can differ significantly from the cooling flow prediction (red dashed line).
}
\end{figure}

Fig. \ref{fig:eng_cool} shows the snapshots of the cooling time (scaled to the shortest cooling time, $\approx 0.022$ Myr) and the temperature PDFs for the simulations with widely varying `engineered' cooling functions. For each $\Lambda$, the emissivity PDF (green solid line) peaks at a temperature ($T$) with the shortest isobaric cooling time. The cooling time panels show that the volume of gas with the shortest cooling time is small because the gas cools quickly over that temperature range. The runs with `engineered' cooling functions show that the volume PDFs as functions of the physical and average temperatures differ substantially, unlike in the fiducial run (shown in Fig. 3 of the Letter). Cooling time is shortest for different physical temperatures for different $\Lambda$s. However, the average cooling-time profiles (blue lines in $t_{\rm cool}$ snapshots) peak where the average temperature is close to the cooling peak, with comparable widths (in units of $\Delta u t_{\rm cool, min}$). A careful study of such engineered cooling functions can lead to important insights into the physics of TRMLs.

%======================
\bibliographystyle{apsrev4-2}
\bibliography{trml_refs}